\pdfoutput=1
\documentclass[%
 reprint,
groupedaddress,
unsortedaddress,
nobibnotes,
amsmath,amssymb,
superscriptaddress,
aps,
prb,
dvipsnames,table,xcdraw
]{revtex4-2}
\usepackage{caption}
\usepackage{subcaption}
\usepackage{graphicx}
\usepackage{dcolumn}
\usepackage{bm}
\usepackage{tikz}
\usetikzlibrary{quantikz}
\usetikzlibrary{shapes.geometric}

\usepackage[backref=none, breaklinks=True]{hyperref} 
\usepackage{color}
\usepackage[framemethod=TikZ]{mdframed}
\usepackage{lipsum}
\usepackage{tabularx}
\usepackage{enumerate}
\usepackage{enumitem}

\newcommand\myshade{85}
\colorlet{mylinkcolor}{violet}
\colorlet{mycitecolor}{YellowOrange}
\colorlet{myurlcolor}{Aquamarine}
\hypersetup{
  linkcolor  = mylinkcolor!\myshade!black,
  citecolor  = mycitecolor!\myshade!black,
  urlcolor   = myurlcolor!\myshade!black,
  colorlinks = true,
}

\usepackage{changes}
\definechangesauthor[name={Santosh}, color=orange]{San}

\newsavebox{\boxRX}
\newsavebox{\boxIQP}
\newsavebox{\boxQAOA}
\begin{document}

\preprint{APS/123-QED}

\title{Quantum-Classical Multiple Kernel Learning}

\author{Ara Ghukasyan}
\affiliation{Agnostiq Inc., 325 Front St W, Toronto, ON M5V 2Y1}
\affiliation{University of Waterloo, 200 University Ave W, Waterloo, ON N2L 3G1}

\author{Jack S. Baker}
\affiliation{Agnostiq Inc., 325 Front St W, Toronto, ON M5V 2Y1}

\author{Oktay Goktas}
\affiliation{Agnostiq Inc., 325 Front St W, Toronto, ON M5V 2Y1}

\author{Juan Carrasquilla}
\affiliation{University of Waterloo, 200 University Ave W, Waterloo, ON N2L 3G1}
\affiliation{Vector Institute,  661 University Ave Suite 710, Toronto, ON M5G 1M1}

\author{Santosh Kumar Radha}
\email{research@agnostiq.ai}
\affiliation{Agnostiq Inc., 325 Front St W, Toronto, ON M5V 2Y1}

\date{\today}

\begin{abstract}
As quantum computers become increasingly practical, so does the prospect of using quantum computation to improve upon traditional algorithms. Kernel methods in machine learning is one area where such improvements could be realized in the near future. Paired with kernel methods like support-vector machines, small and noisy quantum computers can evaluate classically-hard \textit{quantum kernels} that capture unique notions of similarity in data. Taking inspiration from techniques in classical machine learning, this work investigates simulated quantum kernels in the context of \textit{multiple kernel learning} (MKL). We consider pairwise combinations of several classical-classical, quantum-quantum, and quantum-classical kernels in an empirical investigation of their classification performance with support-vector machines. We also introduce a novel approach, which we call \textit{QCC-net} (quantum-classical-convex neural network), for optimizing the weights of base kernels together with any kernel parameters. We show this approach to be effective for enhancing various performance metrics in an MKL setting. Looking at data with an increasing number of features (up to 13 dimensions), we find parameter training to be important for successfully weighting kernels in some combinations. Using the optimal kernel weights as indicators of relative utility, we find growing contributions from trainable quantum kernels in quantum-classical kernel combinations as the number of features increases. We observe the opposite trend for combinations containing simpler, non-parametric quantum kernels.
\end{abstract}

\maketitle
\section{Introduction}
Research towards developing novel computational paradigms aims to overcome the limitations of classical computers. Quantum computing leverages the unique properties of quantum mechanics as a promising alternative for tackling some classically infeasible problems. New algorithms for noisy intermediate-scale quantum (NISQ) devices \cite{Preskill2018} continue to emerge, spurred on by recent demonstrations of quantum supremacy \cite{arute2019quantum, madsen2022quantum}. In the field of machine learning, kernel methods \cite{ShaweTaylor2004} are one promising application of this technology \cite{Biamonte2017}.

Kernel methods, such as support-vector machines (SVMs) \cite{cortes1995support}, are prominent in classical machine learning due to theoretical guarantees associated with their convex loss landscapes. SVMs and related algorithms can perform non-linear classification to capture complex patterns in data. One limitation of kernel methods is that they require computing an $M \times M$ \textit{kernel matrix}, where $M$ is the number of training samples. On the other hand, many quantum kernel methods \cite{Mengoni2019} are immediately suitable for NISQ devices. 
\begin{figure}
    \centering
    \includegraphics[width=0.8\linewidth]{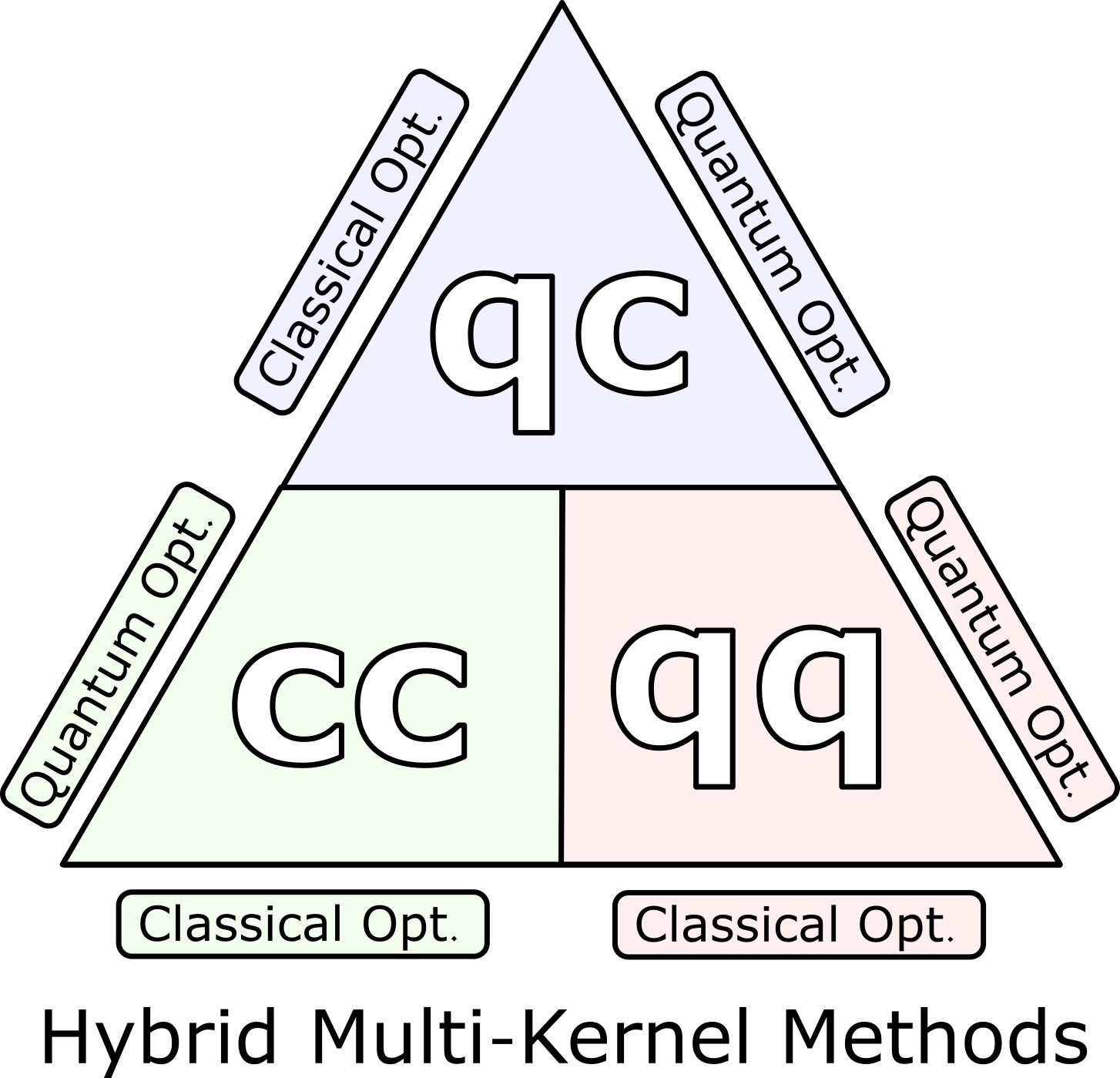}
    \caption{An overview of prospective multiple kernel learning methods in the era of quantum computing, emphasizing combinations of quantum (q) and/or classical (c) kernels, as well as quantum or classical optimization approaches.%
}\label{fig:kernelcomb}
\end{figure}

Quantum kernel methods, like quantum support-vector machines (QSVMs) \cite{rebentrost2014quantum}, have shown promise in various applications \cite{Liu2021}, including supernova classification in cosmology \cite{peters2021machine}, probing phase transitions in quantum many-body physics \cite{sancho2022quantum}, and detecting fraud in finance \cite{kyriienko2022unsupervised}. The crux of the QSVM is embedding input data into quantum states, which allows kernels to be estimated from the overlap of these states. Importantly, the quantum feature spaces accessed by quantum kernel methods can be tailored to data \cite{radha2022generalized,Schuld2019,Hubregtsen2022} by training parametric embeddings.

Multiple kernel learning (MKL) aims to enhance performance by combining different kernels into a single, more expressive kernel \cite{Gonen2011}, in order to learn a wider variety of decision functions. Here, tuning combinations in a data-driven way \cite{Aiolli2015} provides an additional means tailoring kernels. MKL can facilitate feature selection and the incorporation of domain-specific knowledge, which are essential for addressing real-world problems \cite{rakotomamonjy2008simplemkl}. MKL also encourages sparsity in the combination of kernels, effectively performing model selection in learning the optimal weights for component kernels. This can reduce the risk of overfitting and improve generalization performance \cite{suzuki2012fast}. To date, MKL has proven valuable in various applications, including image classification, natural language processing, bioinformatics, and drug discovery \cite{bach2004multiple, yang2011multiple, bach2012optimization}.

As the NISQ era progresses, MKL may prove useful for near term applications that rely on both quantum and classical computing. These paradigms can be combined in six different ways, as illustrated in Fig. \ref{fig:kernelcomb}; utilizing classical-classical (c.c.), quantum-classical (q.c.), or quantum-quantum (q.q.) kernel combinations. Either classical or quantum techniques can also be applied to the subsequent optimization problem as well. For example, recent research has focused on using quantum annealers to solve the SVM problem for fully classical kernels \cite{willsch2020support}. Multiple quantum kernels have also been combined in the fully-quantum (q.q.) setting, using deterministic quantum computing with one qubit (DQC1) \cite{Vedaie2020}, and solving the SVM problem on a classical computer.

A foremost goal of this paper is to systematically address c.c., q.q., and especially q.c. kernel combinations, of which the latter is noticeably lacking in existing literature. Using the classical \textit{EasyMKL} algorithm \cite{Aiolli2015} to optimize kernel weights, we consider pairwise combinations within a set of three quantum and three classical kernels. We also introduce another training step to fine-tune any parametric component kernels in our end-to-end learnable quantum-classical-convex neural network (\textit{QCC-net}) method (also used for time-series analysis in Ref.  \cite{Baker2023}). We report performance metrics for the various combinations in a supervised classification setting using SVMs, across synthetic datasets with a range of two to thirteen features.

The remainder of this paper is organized as follows: Sec. \ref{sec:kernel-methods} provides a brief overview of kernel methods, introduces the concepts behind quantum embedding kernels, and provides additional motivation for hybrid paradigms in MKL. We then present our experimental methodology in Sec. \ref{sec:methodology}, including implementation details, descriptions of the base classical and quantum kernels, as well as details of data preparation and the \textit{QCC-net}. Empirical results are discussed in Sec. \ref{sec:results} in terms of performance metrics and optimal kernel weights. Finally, we draw insights from these results and suggest future directions in Sec. \ref{sec:conclusion}.

\section{Kernel methods \label{sec:kernel-methods}}
Kernel methods are machine learning algorithms that rely on a \textit{kernel function} to compute numerical notions of pairwise similarity. Any function $k(\bm{x}, \bm{x}^{\prime})$ is a kernel if there exists a feature map $\bm{\Phi} : \mathbb{R}^d \rightarrow \mathcal{F}$, such that

\begin{align}
    k(\bm{x}, \bm{x}^{\prime}) = \langle \bm{\Phi}(\bm{x}^{\prime}), \bm{\Phi}(\bm{x})\rangle
\label{eq:inner-product-kernel}
\end{align}
for all $\bm{x} \in X \subset \mathbb{R}^d$.

Kernels provide ``shortcuts'' into a feature space $\mathcal{F}$ via the \textit{kernel trick}.
Consider a linear classifier
\begin{align}
    y_\text{pred}(\bm{x}) = \text{sgn}(\langle \bm{w}, \bm{x} \rangle + b),
\label{eq:linear-decision}
\end{align}
which predicts binary labels $y_i \in \{-1, +1\}$ for points $\bm{x}\in\mathbb{R}^d$, using a separating hyperplane defined by the normal vector $\bm{w} \in \mathbb{R}^d$ and a distance to the origin $b \in \mathbb{R}$. If the classes in $X$ are \textit{not} linearly separable, the algorithm can be attempted in an alternate Hilbert space, finding $\tilde{\bm{w}} \in \mathcal{F}$ and $\tilde{b} \in \mathbb{C}$ such that
\begin{align}
    y_\text{pred}(\bm{x}) = \text{sgn}(\langle \tilde{\bm{w}}, \bm{\Phi}(\bm{x}) \rangle + \tilde{b}).
\label{eq:linear-decision-phi}
\end{align}
correctly classifies $\bm{\Phi}(\bm{x})$ (and thereby $\bm{x}$).

Instead of evaluating Eq. (\ref{eq:linear-decision-phi}) directly, the \textit{representer theorem} \cite{Scholkopf2002} is applied to re-express $\tilde{\bm{w}}$ as a finite linear combination with real coefficients $\alpha_m$, 
\begin{align}
    \tilde{\bm{w}} = \sum_m \alpha_m \bm{\Phi}(\bm{x}^{(m)}),
\label{eq:representer}
\end{align}
which yields a classifier that operates on $\bm{\Phi}(\bm{x})\in\mathcal{F}$, without directly evaluating any inner products:
\begin{align}
    y_\text{pred}(\bm{x}) = \text{sgn}\Bigg(\sum_m \alpha_m k(\bm{x}, \bm{x}^{(m)}) + b\Bigg)
\label{eq:linear-decision-representer}
\end{align}
Thus, the kernel trick enables \textit{non}-linear classification with linear algorithms, using an \textit{implicit} transformation of $X$ through a feature map $\bm{\Phi}$. 

Kernel methods are more transparent to formal analysis (as compared to neural networks, for example) and often lead to convex problems that benefit from provable performance guarantees \cite{Scholkopf2002}. Their connection to supervised \textit{quantum machine learning} (QML) models \cite{Schuld2018}, which can be re-formulated as kernel methods \cite{Schuld2021}, has stoked additional interest for near-term quantum applications. As a more distant prospect, kernel-based problems are especially well suited for fault-tolerant quantum computers \cite{Harrow2009}. 

\begin{table}[b]
\def\arraystretch{1.5}
\begin{tabularx}{\linewidth}{c >{\centering\arraybackslash}X c}
\hline
\textbf{kernel name} & \textbf{kernel function} & \textbf{parameters} \\ 
\hline
Linear & $\langle \bm{x}, \bm{x}^{\prime} \rangle$ & none\\
Polynomial & $(\theta_0 \langle \bm{x}, \bm{x}^{\prime} \rangle + \theta_1)^3$ & $\{\theta_0, \theta_1$\}\\
RBF & $\exp(-\theta_2 ||\bm{x} - \bm{x}^{\prime} ||^2)$ & $\{\theta_2$\}\\
\hline
\end{tabularx}
\caption{Classical kernels considered in this work. Default parameter values are chosen to be $(\theta_0, \theta_1) = (1/d, 1)$ and $\theta_2 = 1$.}
\label{table:classical-kernels}
\end{table}

\savebox{\boxRX}{
\begin{quantikz}[column sep=0.3cm]
    \lstick{} & \gate{R_x(x_1)} & \qw \\
    \lstick{} & \gate{R_x(x_2)} & \qw \\
    \lstick{} & \gate{R_x(x_3)} & \qw \\
\end{quantikz}
}
\savebox{\boxIQP}{
\begin{quantikz}[transparent, column sep=0.3cm]
    \lstick{} & \gate{H} & \gate{R_z(x_1)} & \gate[2]{ZZ_{12} (x_1 x_2)} & \gate[3, label style={yshift=0.7cm}]{ZZ_{13}(x_1 x_3)} & \qw & \qw \\
    \lstick{} & \gate{H} & \gate{R_z(x_2)} & & \linethrough & \gate[2]{ZZ_{23} (x_2 x_3)} & \qw \\
    \lstick{} & \gate{H} & \gate{R_z(x_3)} & \qw & & & \qw \\
\end{quantikz}
}
\savebox{\boxQAOA}{
\begin{quantikz}[transparent, column sep=0.3cm]
    \lstick{} & \gate{R_x(x_1)} & \gate[2]{ZZ_{12} (\theta_1)} & \gate[3, label style={yshift=0.7cm}]{ZZ_{13}(\theta_2)} & \qw & \gate{R_y(\theta_4)} & \qw \\
    \lstick{} & \gate{R_x(x_2)} & & \linethrough & \gate[2]{ZZ_{23} (\theta_3)} & \gate{R_y(\theta_5)} & \qw \\
    \lstick{} & \gate{R_x(x_3)} & \qw & & & \gate{R_y(\theta_6)} & \qw \\
\end{quantikz}
}
\begin{table*}
\centering
\def\arraystretch{1.5}
\begin{tabularx}{\linewidth}{
>{\centering\arraybackslash}X
>{\centering\arraybackslash}X
>{\centering\arraybackslash}X
>{\centering\arraybackslash}X
>{\centering\arraybackslash}c
>{\centering\arraybackslash}X
}
\hline
\textbf{kernel name} & \textbf{embedding unitary} & \textbf{Eqs.} & \textbf{parameters} & \textbf{embedding circuit} $(U_{\bm{\theta}}(\bm{x}))$ & \textbf{ref.} \\ 
\hline
& & & & \\
RX & $\mathcal{R}_{\text{X}}(\bm{x})$ & (\ref{eq:R_a}) & - & \resizebox{!}{1cm}{\usebox\boxRX} & - \\
\hline
& & & & \\
IQP & $V(\bm{x}) H^{\otimes N}$ & (\ref{eq:IQP-equation}) & - & \resizebox{!}{1.5cm}{\usebox\boxIQP} & \cite{Havlivcek2019}  \\
\hline
& & & & \\
QAOA & $W(\bm{\theta}) \mathcal{R}_{\text{X}}(\bm{x})$ & (\ref{eq:R_a}, \ref{eq:qaoa-w}) & $\bm{\theta} \in \mathbb{R}^{2N}$ & \resizebox{!}{1.4cm}{\usebox\boxQAOA} & \cite{Farhi2014, Lloyd2020quantum} \\
\hline
\end{tabularx}
\caption{Quantum kernels considered in this work. Circuit diagrams provide a 3-qubit example of the \textit{embedding} circuit (\textit{i.e.} one half of the kernel circuit). Initial parameters for the QAOA circuit are chosen uniformly at random from $[0,\,2\pi]^{2N}$. Equations defining the embedding circuits are provided in Sec. \ref{subsection:clasical-and-quantum-kernel-selection}.}
\label{table:quantum-kernels}
\end{table*}

\subsection{Quantum Kernels}
Quantum embedding kernels (QEKs) \cite{Schuld2019, Hubregtsen2022} implement a similarity measures via fidelity estimates between data-representing quantum states. QEKs are an attractive prospect insofar as classically-hard kernel functions can be realized with parametric operations that encode data points as quantum states \cite{Havlivcek2019,Liu2021}. To achieve efficient and expressive  QEKs, the quantum feature space should be readily parametrized by only a handful of gate variables (here denoted by $\bm{\theta}$).

To construct a QEK, one must choose a unitary transformation $U_{\bm{\theta}}$ to define the embedding,
\begin{align}
    \ket{\bm{\Phi}(\bm{x})} = U_{\bm{\theta}}(\bm{x})\ket{0}.
\label{eq:quantum-embedding}
\end{align}
Here, a corresponding feature map is one that takes $\bm{x}$ to a \textit{density matrix} $\rho(\bm{x}) = \ket{\bm{\Phi}(\bm{x})}\bra{\bm{\Phi}(\bm{x})}$. Note that any possibility for quantum advantage is lost if the embedding is too simple \cite{Havlivcek2019}, so $U$ must be chosen carefully. The QEK itself is defined by the Frobenius inner product $\langle \rho(\bm{x}^{\prime}), \rho(\bm{x}) \rangle$, or equivalently
\begin{align}
    \text{Tr} \{ \rho(\bm{x}^{\prime}) \rho(\bm{x}) \} = | \braket{\bm{\Phi}(\bm{x}^{\prime})}{\bm{\Phi}(\bm{x})} |^2 = k_{\bm{\theta}}(\bm{x}, \bm{x}^{\prime}).
\label{eq:quantum-kernel}
\end{align}
Any of several existing methods for fidelity estimation \cite{Cincio2018, Fanizza2020, Huang2020} can be used to evaluate Eq. (\ref{eq:quantum-kernel}) in practice, allowing a hybrid classifier to be implemented by calling $k_{\bm{\theta}}(\bm{x}, \bm{x}^{\prime})$ inside a classical linear algorithm (e.g. Eq. (\ref{eq:linear-decision-representer})). More generally, the modular nature of kernel methods provides a convenient setting for high-level manipulation of both quantum and/or classical kernels.

While the kernel trick is simple to apply, it remains challenging to determine an effective kernel for a given classification problem. To this end, heuristic \cite{Hubregtsen2022} or cost-based optimization \cite{Havlivcek2019} can be leveraged to improve the performance of parametric kernels. MKL is a complementary approach that considers different kernels in combination, and \textit{determines} the weights for component kernels.
 
\subsection{Multiple Kernel Learning}
The goal in MKL is to improve a kernel method's performance by introducing a novel notion of similarity derived from multiple distinct kernels. Typical use cases for MKL include affecting feature selection \cite{Brouard2022, Xue2020}, enabling anomaly detection \cite{Gautam2019}, and enhancing expressivity \cite{Vedaie2020}. Many kernel combination strategies have been proposed for MKL \cite{Gonen2011}, although weighted linear combinations are often effective \cite{Aiolli2015,Vedaie2020}.

A combination function $f_{\bm{\gamma}} : \mathbb{R}^R \rightarrow \mathbb{R}$ and a base set of kernels $k_{\bm{\theta}}^{(r)} : \mathbb{R}^d \times \mathbb{R}^d \rightarrow \mathbb{R}$ define a combined kernel \cite{Gonen2011}
\begin{align}
    k_{\bm{\theta},\bm{\gamma}}(\bm{x}, \bm{x}^{\prime}) := f_{\bm{\gamma}}\Big(\{k_{\bm{\theta}}^{(r)}(\bm{x}, \bm{x}^{\prime})\}_{r=1}^R\Big)
\label{eq:combined-kernel}
\end{align}
with adjustable parameters $\bm{\theta}$ and $\bm{\gamma}$ for $k_{\bm{\theta}}^{(r)}$ and $f_{\bm{\gamma}}$, respectively. For brevity of notation, $\bm{\theta}$ is assumed to contain all kernel parameters and each kernel $k_{\bm{\theta}}^{(r)}$ is assumed to utilize only a subset ``$r$'' of $\bm{\theta}$.

To be a valid kernel, the resulting $k_{\bm{\theta}, \bm{\gamma}}$ must be \textit{positive semi-definite}, which means the inequality
\begin{align}
    \bm{z}^T K_{\bm{\theta},\bm{\gamma}} \bm{z} \geq 0
\label{eq:positive-semi-definite}
\end{align}
must hold for all $\bm{z} \in \mathbb{R}^M$, where $M = \#\{\hat{Z}\}$ for any (improper) subset $\hat{Z}$ of the \textit{training} set, $\hat{X} \subset X$. Here, $K_{\bm{\theta},\bm{\gamma}} \in \mathbb{R}^{M \times M}$ is the Gram matrix \cite{Scholkopf2002} of all pairwise kernels, $k_{\bm{\theta},\bm{\gamma}}(\bm{x}, \bm{x}^\prime)$, for $\bm{x},\bm{x}^\prime \in \hat{Z}$. If the Gram matrices of all $R$  base kernels satisfy Eq. (\ref{eq:positive-semi-definite}), then their additive or multiplicative combinations are also guaranteed to be valid kernels \cite{Steinwart2008}.

\begin{figure}[t!]
    \centering
    \includegraphics[width=0.9\linewidth]{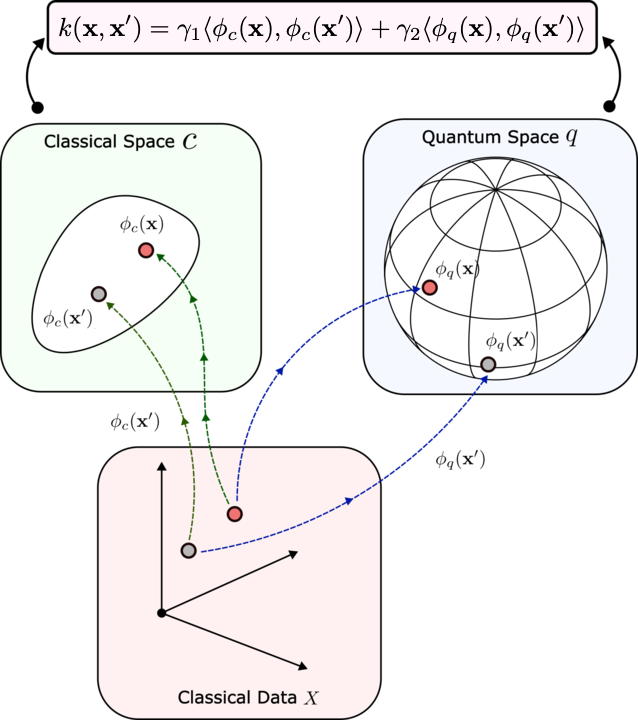}
    \caption{Illustration of a hybrid kernel that combines inner products in $C_1$, a generic (``classical'') Hilbert space, with inner products from $Q_1$, the state space of a quantum system.}
    \label{fig:hybrid-kernel}
\end{figure}

A convex linear combination of base kernels (Fig. \ref{fig:hybrid-kernel}) is constructed by taking
\begin{align}
    f_{\bm{\gamma}} = \langle \bm{\gamma}, \cdot \rangle
\label{eq:f-gamma-linear}
\end{align}
with the following constraints on $\bm{\gamma}$:
\begin{align}
    ||\bm{\gamma}||_1 = 1,\quad \gamma_r \geq 0.
\label{eq:gamma-constraint}
\end{align}
As a weighted average, this form is both easy to interpret and computationally convenient. Using Eq. (\ref{eq:inner-product-kernel}), we can infer that the resulting kernel,
\begin{align}
    k_{\bm{\theta}, \bm{\gamma}}(\bm{x}, \bm{x}^{\prime}) &:= \sum_{r=1}^R \bm{\gamma}_r k^{(r)}_{\bm{\theta}}(\bm{x}, \bm{x}^{\prime}),
\label{eq:kernel-linear-combination}
\end{align}
corresponds to a feature map
\begin{align}
    \bm{\Phi}_{\bm{\theta},\bm{\gamma}}(\bm{x}) = [\sqrt{\gamma_1}\bm{\Phi}_{\bm{\theta}}^{(1)}(\bm{x}), ... , \sqrt{\gamma_R}\bm{\Phi}_{\bm{\theta}}^{(R)}(\bm{x})]^T.
    \label{eq:combined-feature-map}
\end{align}
Here, data-driven optimization of $\bm{\gamma}$ effectively decides the importance of each component feature map. Once all the relevant kernels have been evaluated, the MKL technique does not distinguish between kernels of quantum and/or classical nature. Instead, the kernel values are simply weighted to maximize an optimization objective (Sec. \ref{EasyMKL}) in view of the training data.  

To contrast, the q.q. implementation by Vedaie \textit{et al.} \cite{Vedaie2020} involves preparing a parametrized \textit{mixed state},
\begin{align}
    \rho_{\bm{\gamma}} = \sum_{r = 1}^{R} \gamma_{r} \ket{r}^{\otimes q}\bra{r}^{\otimes q},
\label{vedaie-mixed-state}
\end{align}
which represents an $R$-component kernel of the form in Eq. (\ref{eq:kernel-linear-combination}); weighted by the classical probabilities of the mixture, and computed via the expectation value of an encoding operator. Here, each individual kernel is evaluated at the same time, using a single quantum circuit.

In our approach, the kernel (or Gram) matrices, 
\begin{align}
    [K^{(r)}_{\bm{\theta}}]_{\bm{x},\bm{x}^\prime} = k^{(r)}_{\bm{\theta}}(\bm{x}, \bm{x}^{\prime}),\quad\forall\bm{x},\bm{x}^\prime\in\hat{X}
\label{eq:gram-matrix}
\end{align}
are computed separately for each quantum or classical kernel. The combination parameters ($\bm{\gamma}_r$) are then determined classically, using these matrices and the training labels as inputs. This allows us to treat c.c., q.q., and q.c. kernel combinations identically, within a general optimization framework.

\subsubsection{Quantum-Classical Kernel Combinations}
\begin{figure}[h!]
    \centering
    \includegraphics[width=0.85\linewidth]{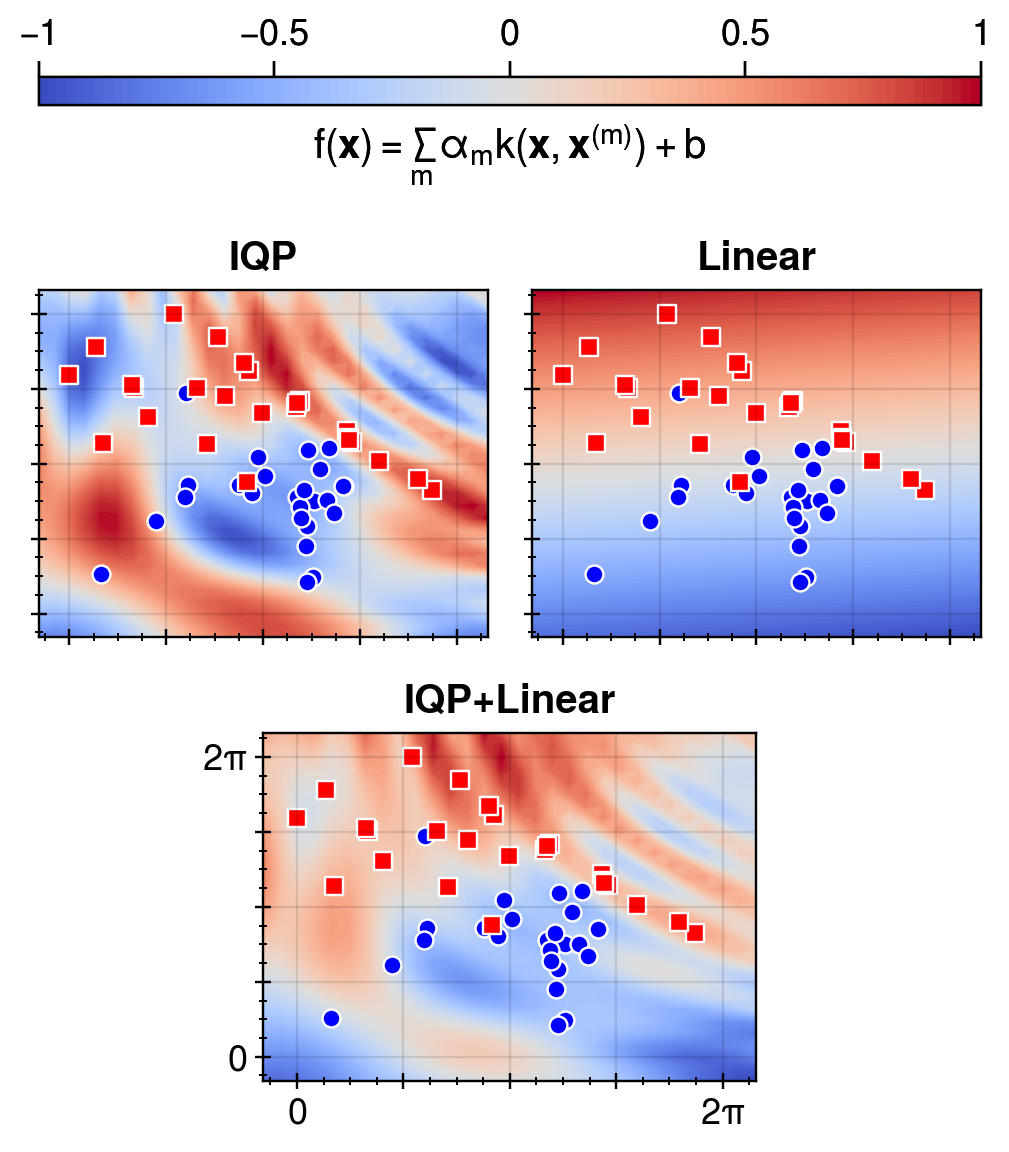}
    \caption{Examples of trained SVM decision functions from the IQP (Tab. \ref{table:quantum-kernels}) and Linear kernels (Tab. \ref{table:classical-kernels}). Top panels show each kernel individually and the bottom panel shows their linear combination, with weights $\gamma_1 = 0.4$ and $\gamma_2 = 0.6$ for the IQP and Linear kernels, respectively. The pictured training set represents a typical ($d=2$) instance of the datasets considered in this study. All horizontal and vertical axes range from 0 to $2\pi$.}
    \label{fig:decision-interpolation}
\end{figure}

Since computing common classical kernels adds no significant overhead compared to querying quantum computers or simulating quantum circuits, a useful q.c. combination could be an easy way to make the most of available NISQ hardware. Feature spaces associated with quantum kernels are unique insofar as (1) they can be classically-hard to simulate, and (2) are directly modelled by physical quantum states. The first point here is necessary for achieving \textit{quantum advantage} \cite{Liu2021} (though this does not \textit{per se} guarantee better learning performance), while the second point means that the feature space is \textit{directly} tunable in an efficient manner, unlike useful feature spaces in the traditional setting. Using an MKL strategy like Eq. (\ref{eq:kernel-linear-combination}) creates a weighted concatenation of feature vectors, which leads to an overall kernel that interpolates between component kernels (Fig. \ref{fig:hybrid-kernel}). Adding a classical kernel to a complicated (and often periodic) quantum kernel also introduces a natural means of trainable regularization (Fig. \ref{fig:decision-interpolation}), which can improve model performance beyond the training set. This effect is adjustable and could be eliminated entirely if the classical kernel is indeed superfluous. On the other hand, the optimal kernel weights could reveal that the quantum kernel is actually ineffective for the problem at hand.

\subsubsection{Kernel Training and MKL Optimization}
The process of training the kernel parameters ($\bm{\theta}$) typically represents a separate consideration with respect to optimizing the kernel combination weights ($\bm{\gamma}$). In this work, kernel parameter training relies on a stochastic, gradient-based method \cite{kingma2014adam}, while weights optimization involves solving a convex quadratic problem based on the \textit{EasyMKL} algorithm \cite{Aiolli2015}. The \textit{QCC-net} introduced in Sec. \ref{EasyMKL} combines these processes in a fully differentiable way, with the MKL objective substituted as the training loss. As we shall demonstrate in Sec. \ref{subsec:training-impact}, parameter training is necessary in some cases for the MKL algorithm to distinguish component kernels.

\section{Methodology \label{sec:methodology}}
\subsection{Design \label{subsec:design}}
To explore c.c., q.c., and q.q. MKL methods side by side, we considered binary classification tasks using all \textit{pairwise} combinations from a set of three classical (Tab. \ref{table:classical-kernels}) and three quantum (Tab. \ref{table:quantum-kernels}) kernels. To quantify classification performance and characterize the kernels, we used the metrics in Tab. \ref{table:metrics}. Here, the base metrics $(T/F)P$ and $(T/F)N$ correspond to the true/false positive and true/false negative counts, respectively. These are used to compute the classification metrics in the lower half of Tab. \ref{table:metrics} according to the corresponding definition.

\begin{table}[b]
\def\arraystretch{1.5}
\begin{tabularx}{\linewidth}{c >{\centering\arraybackslash}X}
\hline
\textbf{metric} & \textbf{definition} \\
\hline
TP & $\#\{\bar{\bm{x}}\,\,\text{s.t.}\,\,y_{\text{pred}}(\bar{\bm{x}}) = \bar{y} = 1\}$ \\[1.25ex]
TN & $\#\{\bar{\bm{x}}\,\,\text{s.t.}\,\,y_{\text{pred}}(\bar{\bm{x}}) = \bar{y} = -1\}$ \\[1.25ex]
FP & $\#\{\bar{\bm{x}}\,\,\text{s.t.}\,\,y_{\text{pred}}(\bar{\bm{x}}) = 1, \bar{y} = -1\}$ \\[1.25ex]
FN & $\#\{\bar{\bm{x}}\,\,\text{s.t.}\,\,y_{\text{pred}}(\bar{\bm{x}}) = -1, \bar{y} = 1\}$ \\[1.25ex]
\hline
Accuracy & $\frac{TP + TN}{TP + TN + FP + FN}$ \\[2ex]
AUCROC & area under $(\frac{TP}{TP + FN}, \frac{FP}{FP + TN})\Big|_t$ \\[2ex]
Margin & $\min\{||\bm{\Phi}_{\bm{\theta}, \bm{\gamma}}(\bm{x}) - \bm{\Phi}_{\bm{\theta}, \bm{\gamma}}(\bm{x}^{\prime})||\,\,\text{s.t.}\,\,y \neq y^\prime\}$ \\[2ex]
Spectral Ratio & $\sum[K_{\bm{\theta},\bm{\gamma}}]_{ii} / \sqrt{\sum[K_{\bm{\theta},\bm{\gamma}}]_{ij}^2}$ \\[2ex]
\hline
\end{tabularx}
\caption{Base metrics (top four), classification metrics (accuracy, AUCROC), and kernel metrics (margin, spectral ratio)  used for combined comparisons in Sec. \ref{sec:results}. Classification metrics are computed from testing outcomes and kernel metrics are computed from training outcomes.}
\label{table:metrics}
\end{table}

The reported accuracy is the number of correct predictions divided by total number of predictions. The area under the receiver operator characteristic (AUCROC) is computed across true/false positive \textit{rates} over decreasing thresholds, $t$. These thresholds are obtained from $y_\text{pred}(\bar{\bm{x}})$ and $\bar{y}$ for all $\bar{\bm{x}}$ in the \textit{testing} set $\bar{X} = X \setminus \hat{X}$, using the \texttt{roc\_auc\_score} function from \cite{scikit-learn}.

The final two metrics in Tab. \ref{table:metrics}, the margin and spectral ratio, are computed on the Gram matrix that represents the ``kernelized'' training data. These metrics characterize the kernel function and its embedding, unlike the accuracy and AUCROC, which are measures of \textit{classification} performance. The margin is the smallest distance between points in feature space ($\mathcal{F}$) which belong to different classes. It is the maximization target for the SVM algorithm \cite{Steinwart2008} and has a fixed upper bound for a given dataset $\hat{X}$ and feature map $\bm{\Phi}$.

The spectral ratio is the sum of the diagonal elements of $K_{\bm{\theta},\bm{\gamma}}$ divided by its Frobenius (or $L_{2,2}$) norm. For \textit{bounded} kernels (\textit{i.e.}, RBF and all quantum kernels) the diagonal sum is always equal to $\#\{\hat{X}\}$ (i.e. the size of $\hat{X}$), because these kernels evaluate to unity on identical input pairs. Since off-diagonal terms correspond to non-identical input pairs, the maximum spectral ratio of 1 is obtained with a kernel
\begin{align}
   k_{\bm{\theta},\bm{\gamma}}(\bm{x},\bm{x}^\prime) = \delta_{\bm{x},\bm{x}^\prime},  
\end{align}
which is a Kronecker delta function on the domain $\bm{x},\bm{x}^\prime \in \hat{X}$. This limit corresponds to poor classification in general, as only identical points are considered ``similar''. Conversely, the minimum spectral ratio $1/\#\{\hat{X}\}$ is obtained with the constant function
\begin{align}
    k_{\bm{\theta},\bm{\gamma}}(\bm{x},\bm{x}^\prime) = 1,
\end{align}
which considers every pair of points to be maximally similar. Therefore, a reasonable spectral ratio corresponds to some value between these extremes.

For combinations containing \textit{unbounded} kernels (Linear, Polynomial), we normalize the component Gram matrix (Eq. (\ref{eq:gram-matrix})) before the sums over $K_{\bm{\theta},\bm{\gamma}}$ (Tab. \ref{table:metrics}) are computed. Normalization additionally helps to stabilize the kernel weighting algorithm (Sec. \ref{EasyMKL}) in general. However, this results in $\sum [K_{\bm{\theta}}^{(r)}]_{ii} < \#\{\hat{X}\}$, which (unless $\gamma_r$ is zero) unnaturally lowers of the \textit{combined} kernel's spectral ratio. Direct comparisons of spectral ratios are therefore not reliable between combinations that contain only bounded kernels versus those containing one or more unbounded kernels. Comparisons \textit{within} these two groups, however, are fully justified.

\subsubsection{Classical and Quantum Kernel Selection \label{subsection:clasical-and-quantum-kernel-selection}}
Among the classical set, the Linear kernel corresponds to a feature map whose feature space is (trivially) the native data space ($\mathbb{R}^d$). Any prediction function learned with the Linear kernel is therefore strictly linear itself. We also consider a cubic Polynomial kernel and the radial basis function (RBF) kernel as more ``powerful'' examples of common classical kernels. Both of the latter two are parametric and produce non-trivial feature maps. Any feature space of the RBF kernel is in fact infinite-dimensional \cite{Steinwart2006}, and the kernel itself is known to be universal \cite{Micchelli2006}.

On the quantum side, we construct a QEK from each embedding unitary in Tab. \ref{table:quantum-kernels} as per Eqs. (\ref{eq:quantum-embedding},\ref{eq:quantum-kernel}). The shorthand
\begin{align}
    \mathcal{R}_{\text{A}} (\bm{z}) = \exp\Bigg(-\frac{i}{2} \sum_{p=1}^{m} z_p \text{A}_p \Bigg)
    \label{eq:R_a}
\end{align}
is introduced here to represent single-qubit $\text{A}$-rotations, where $\text{A}_p$ are Pauli $X$, $Y$, or $Z$ gates acting on qubit $p$ of an $N$-qubit circuit; taking the rotation angle from component $p \leq m \leq N$ of the input vector $\bm{z} \in \mathbb{R}^m$.

The simplest quantum kernel that we consider is RX (Tab. \ref{table:quantum-kernels}), which encodes each component of $\bm{x}\in\mathbb{R}^d$ onto one of $N=d$ qubits using $\mathcal{R}_{\text{X}}(\bm{x})$ as the embedding unitary. The IQP \cite{Havlivcek2019} kernel is defined by the ansatz
\begin{align}
    V(\bm{x}) = \Bigg\{\exp\Bigg(-\frac{i}{2}\sum_{p\neq q} x_p x_q Z_p Z_q \Bigg)\Bigg\} \mathcal{R}_\text{Z}(\bm{x}),
    \label{eq:IQP-equation}
\end{align}
which uses of $\mathcal{R}_\text{Z}$ rotations followed by two-qubit gates on all pairs of qubits to encode data. Lastly, the \textit{variational} ansatz,
\begin{align}
    W(\bm{\theta}) = \mathcal{R}_\text{Y}(\bm{\theta}) \Bigg\{\exp\Bigg(-\frac{i}{2}\sum_{p\neq q} \theta_{pq} Z_p Z_q \Bigg)\Bigg\}.
    \label{eq:qaoa-w}
\end{align}
defines the QAOA \cite{Farhi2014, Lloyd2020quantum} kernel. While similar to IQP, the QAOA kernel features \textit{parametric} transformations after a single set of encoding gates. Regarding all three embedding circuits, we utilize the minimum, single layer ansatz in every case, as shown in Tab. \ref{table:quantum-kernels}. 

\subsubsection{QCC-net Optimization for MKL \label{EasyMKL}}
Given $M$ training samples from $\hat{X} \subset X \subset \mathbb{R}^d$ and their labels as a matrix $\hat{Y} = \text{diag}(y_1, ..., y_M)$, where $y_i \in \{-1, 1\}$, the combination weights are determined to maximize the total distance (in \textit{feature} space) between positive and negative samples. Following \cite{Aiolli2015}, this problem is formulated as
\begin{align}
    \max_{||\bm{\gamma}|| = 1}\, \min_{\bm{\phi}}\, (1 - \lambda) \bm{\gamma}^T \bm{d}_{\bm{\theta}}(\bm{\phi}) + \lambda ||\bm{\phi}||^2_2,
\label{eq:easymkl-minmax-problem}
\end{align}
where the vector of distances, $\bm{d}_{\bm{\theta}}(\bm{\phi}) \in \mathbb{R}^R$, has components
\begin{align}
    d_{\bm{\theta}}^{(r)}(\bm{\phi}) = \bm{\phi}^T\hat{Y} \hat{K}_{\bm{\theta}}^{(r)} \hat{Y} \bm{\phi},
    \label{eq:separation-component}
\end{align}
and the variable $\bm{\phi} \in \mathbb{R}^M$ is subject to
\begin{align}
    \sum_{i|y(\bm{x}) = 1} \bm{\phi}_i = \sum_{i|y(\bm{x}) = -1} \bm{\phi}_i = 1.
\label{eq:phi-constraint}
\end{align}
Considering $0 \leq \lambda < 1$, a solution $\bm{\gamma}^\star$ parallel to $\bm{d}_{\bm{\theta}}(\bm{\phi})$ is evident for the outer maximization problem:
\begin{align}
    \bm{\gamma}^\star = \frac{\bm{d}_{\bm{\theta}}(\bm{\phi})}{||\bm{d}_{\bm{\theta}}(\bm{\phi})||_2}.
\label{eq:gamma-solution}
\end{align}
This reduces Eq. (\ref{eq:easymkl-minmax-problem}) to the (convex) minimization problem,
\begin{align}
    \mathcal{L}_{\bm{\phi}_{\text{min}}}(\bm{\theta}) = \min_{\bm{\phi}}\, (1 - \lambda) || \bm{d}_{\bm{\theta}} (\bm{\phi}) ||_2 + \lambda ||\bm{\phi}||^2_2,
\label{eq:easymkl-problem}
\end{align}
whereby the solution $\bm{\phi}_{\text{min}}$ leads to the optimal kernel weights via Eq. (\ref{eq:gamma-solution}), given fixed values of the kernel parameters $\bm{\theta}$.

We implemented the above algorithm, which describes the process of determining the kernel weights ($\bm{\gamma}^\star$), using the convex program solver CVXPY \cite{cvxpy2016, cvxpy2018} together with the default splitting conic solver (SCS) \cite{odonoghue2016}. In order to integrate optimization of the kernel parameters ($\bm{\theta}$), we expressed the problem in Eq. (\ref{eq:easymkl-problem}) as a differentiable neural network layer using the CVXPY extension \texttt{cvxpylayers} \cite{cvxpylayers2019}, which uses \texttt{diffcp} \cite{diffcp2019} internally. This allowed the gradients
\begin{align}
    \frac{\partial \mathcal{L}_{\bm{\phi}_{\text{min}}}(\bm{\theta})}{\partial \bm{\theta}} := \Bigg(\frac{\partial \mathcal{L}_{\bm{\phi}_{\text{min}}}(\bm{\theta})}{\partial \theta_0},\,...\,,\frac{\partial \mathcal{L}_{\bm{\phi}_{\text{min}}}(\bm{\theta})}{\partial \theta_{N-1}} \Bigg)
\label{eq:qcc-net-gradient}
\end{align}
to be obtained at intermediate steps in the overall computation. Access to these gradients, in turn, allows a gradient-based optimizer to be used for training the kernel parameters to maximize $\mathcal{L}_{\bm{\phi}_{\text{min}}}(\bm{\theta})$ with respect to $\bm{\theta}$. (We utilized the Adam optimizer \cite{kingma2014adam} in this work.) As illustrated in FIG. 1 of Ref. \cite{Baker2023}, the complete \textit{QCC-net} algorithm proceeds as follows:
\begin{enumerate}[label=(\roman*)]
    \item Starting with $R$ base kernels, compute $\hat{K}_{\bm{\theta},\bm{\gamma}}$ assuming balanced $\bm{\gamma}$ (\textit{i.e.} all components equal to $1/R$), and using random/default initial kernel parameters $\bm{\theta}_l = \bm{\theta}$, where $l = 0$ initially.
    \item Solve the cone problem in Eq. (\ref{eq:easymkl-problem}) using CVXPY+SCS to determine $\bm{\phi}_{\text{min}}$ at fixed $\bm{\theta}_l$. If the optimizer's termination criteria are met, advance to (iv). Else, continue to (iii).
    \item Use the gradients in Eq. (\ref{eq:qcc-net-gradient}) to update the kernel the parameters $\bm{\theta}_{l} \rightarrow \bm{\theta}_{l+1}$, maximizing loss function $\mathcal{L}_{\bm{\phi}_{\text{min}}}(\bm{\theta})$ with respect to $\bm{\theta}$, then return to step (i) with $\bm{\theta}_{l} := \bm{\theta}_{l+1}$. 
    \item Set $\bm{\theta}^\star = \bm{\theta}_{l}$, obtain $\bm{\gamma}^\star$ from $\bm{\phi}_{\text{min}}$, and compute the final kernel matrix.
\end{enumerate}

The QML framework Pennylane \cite{pennylane2022} was used to implement and simulate quantum circuits with a PyTorch- \cite{pytorch2019} compatible interface for gradient calculations and GPU acceleration. We also used Covalent (see \footnote{Covalent: \url{https://www.covalent.xyz}}) to facilitate distributed computations.

\subsection{Dataset Preparation and Preprocessing \label{subsec:dataset-preparation}}
We utilized the \texttt{scikit-learn} software package \cite{scikit-learn} to synthesize instances of generic, single-label datasets for binary classification \footnote{\texttt{make\_classification}: \url{https://scikit-learn.org/stable/modules/generated/sklearn.datasets.make_classification.html}} (based the method in \cite{Guyon2003}). Each dataset instance consisted of 100 samples ($\bm{x} \in \mathbb{R}^d$), with $d$ informative features, split 50:50 into training and testing subsets. Samples belonging to each class were distributed evenly between two $d$-variate Gaussian clusters with randomly correlated features; for $d$ in our experiments ranging from 2 and 13. Our initial investigations determined that all kernel combinations performed very well on linearly separable datasets, as expected. Therefore, to reveal meaningful differences between kernel combinations, we used a ``class separation'' of 1.0 (see \footnotemark[1]), ensuring with high probability that the two clusters were \textit{not} linearly separable. All features were scaled (using \footnote{\texttt{MinMaxScaler}: \url{https://scikit-learn.org/stable/modules/generated/sklearn.preprocessing.MinMaxScaler.html}}) into the interval $[0, 2\pi]$ in preprocessing. We also utilized an SVM implementation from \texttt{scikit-learn} \footnote{\texttt{SVC}: \url{https://scikit-learn.org/stable/modules/generated/sklearn.svm.SVC.html}}.

A total 120 instances of this classification problem were computed for each of the three result types (Tab. \ref{table:result-types}), across $d = 2,3,...,13$ features, with 10 repetitions per value of $d$. The Supplementary Information \footnote{See Supplementary Information [publisher url] for examples of two-dimensional datasets.} contains representative examples for the easily visualized $d=2$ case, with one training set also visible in Fig. \ref{fig:decision-interpolation}. 

In the section that follows, performance metrics based on the outcomes of these experiments are used to demonstrate the \textit{QCC-net} optimization approach. We also identify how kernel parameter optimization influences the selection of kernel weights in pairwise combinations of q.c., q.q., and c.c. kernels, and conclude with a brief discussion to reconcile the prior results.

\begin{table}
\def\arraystretch{1.2}
    \begin{tabularx}{\linewidth}{
    c
    >{\centering\arraybackslash}X
    >{\centering\arraybackslash}X
    }
    \hline
    \textbf{result type} & $\bm{\theta}$ & $\bm{\gamma}$ \\
    \hline
    (i) & default/random & uniform\\
    (ii) & default/random & optimized\\
    (iii) & trained & optimized\\
    \hline
    \end{tabularx}
\caption{The three types of results considered in this section. Types are distinguished by the inclusion of kernel parameter ($\bm{\theta}$) training and/or weights ($\bm{\gamma}$) optimization.}
\label{table:result-types}
\end{table}

\section{Results and Discussion\label{sec:results}}
As shown in Tab. \ref{table:result-types}, the results considered in this section are subdivided into three distinct types: (i) \textit{non-optimized}, (ii) \textit{semi-optimized}, and (iii) \textit{fully optimized}. Note that only type (iii) results include kernel parameter training. The other two types of results feature default (Tab. \ref{table:classical-kernels}) or randomly chosen (Tab. \ref{table:quantum-kernels}) parameter values. For type (i) and type (ii) results, default kernel parameters are kept constant throughout, whereas random kernel parameters are uniquely generated for each instance of the problem.

\begin{figure*}[t]
    \centering
    \includegraphics[width=0.85\linewidth]{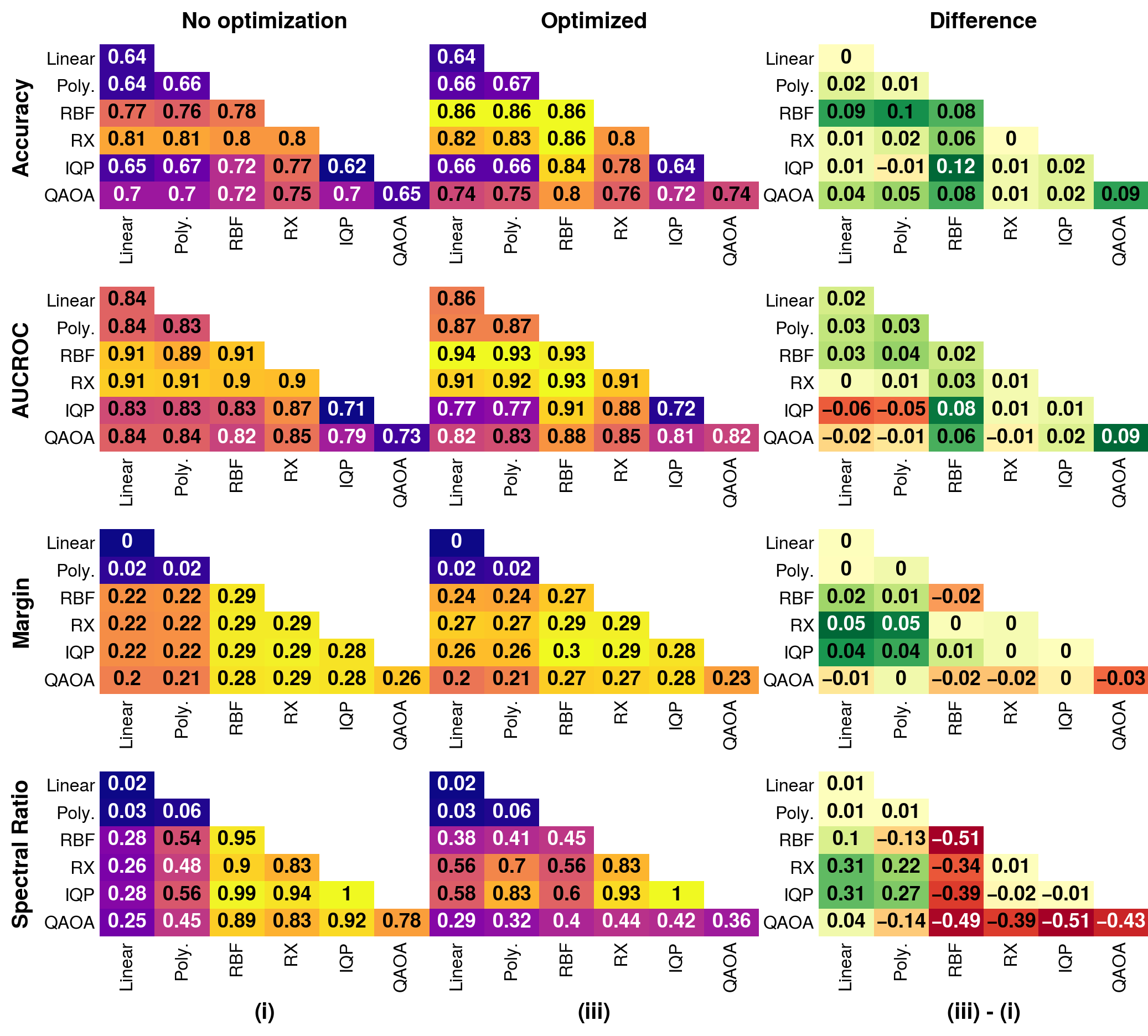}
    \caption{Median kernel metrics over 120 instances, across datasets with $d=2$ to $13$ features, for every unique kernel combination. Colour scales are normalized among adjacent pairs in columns (i) and (iii), which are labelled according to the type of results contained therein. The final column contains the difference, (i) subtracted from (iii). Entries in the difference column are true values rounded to 2-digit precision.}
    \label{fig:metrics_a-r-m-s}
\end{figure*}

\subsection{Performance Comparisons \label{subsec:results-performance}}
With the chosen MKL strategy, a combination of \textit{identical} kernels is equivalent to that specific kernel individually, due to the symmetry present in Eq. (\ref{eq:kernel-linear-combination}). That is, the base kernel is recovered when $k^{(r)}_{\bm{\theta}}(\bm{x}, \bm{x}^{\prime}) \equiv k^{(1)}_{\bm{\theta}}(\bm{x}, \bm{x}^{\prime})$ for all $r$:
\begin{align}
    k_{\bm{\theta}, \bm{\gamma}}(\bm{x}, \bm{x}^{\prime}) &:= k^{(1)}_{\bm{\theta}}(\bm{x}, \bm{x}^{\prime})\sum_{r=1}^R \bm{\gamma}_r = k^{(1)}_{\bm{\theta}}(\bm{x}, \bm{x}^{\prime})
    \label{eq:kernel-combination-collapse}
\end{align}
Hence, the \textit{diagonal} entries in Fig. \ref{fig:metrics_a-r-m-s} indicate the given metric for the lone base kernel. Equivalent values are expected for diagonal entries in columns (i) and (iii) if the base kernel is also non-parametric (Linear, RX, and IQP). In all other cases, however, the kernel combination (or a parametric base kernel) stands to benefit from parameter $(\bm{\theta})$ and/or kernel weights $(\bm{\gamma})$ optimization.

To provide an overall measure of optimization efficacy, we list the total differences between non-optimized (i) and fully optimized (iii) results in Tab. \ref{table:metrcs_a-r-m-s_sum_improvement}. The table is obtained by adding the differences in the final column of Fig. \ref{fig:metrics_a-r-m-s} according to the following scheme: For a given metric, values for kernel combinations $A$-$B$ contribute to totals for both kernel $A$ and kernel $B$, whereas $A$-$A$ combinations contribute only once to the total for kernel $A$.

\subsubsection{Accuracy}
Looking at accuracy outcomes, the RBF- and RX- containing kernels exhibit the largest and second-largest scores, respectively, both with and without optimization. Meanwhile, the RBF- and QAOA- containing kernels exhibit the largest and second-largest \textit{improvements} in accuracy, as compared to the other combinations. The greatest overall improvement in accuracy corresponds to the IQP-RBF kernel combination. Prior to optimization, and indeed after, the IQP-containing kernels exhibit the lowest overall accuracy, considering score totals in the same way as above. This improvement can be attributed to regularization via inclusion of the smooth RBF kernel together with IQP. This point was exemplified by Fig. \ref{fig:decision-interpolation}, earlier in this text. Apart from RBF and QAOA, the Polynomial kernel (Tab. \ref{table:classical-kernels}) is the only other parametric kernel. Compared to the related (and non-parametric) Linear kernel, the Polynomial kernel does not exhibit a significant difference in accuracy scores for the classification problem at hand, even after optimization.

\begin{table*}[t!]
\def\arraystretch{1.25}
    \begin{tabularx}{\linewidth}{
    >{\centering\arraybackslash}X
    >{\centering\arraybackslash}X
    >{\centering\arraybackslash}X
    >{\centering\arraybackslash}X
    >{\centering\arraybackslash}X
    }
    \hline
    & \textbf{Accuracy} & \textbf{AUCROC} & \textbf{Margin} & \textbf{Spectral Ratio} \\ 
    \hline
    Linear & 0.18 & 0.02 & \textbf{0.10} & 0.75 \\
    Polynomial & 0.18 & 0.02 & \textbf{0.10} & 0.22 \\
    RBF & \textbf{0.55} & \textbf{0.26} & 0.02 & -1.76 \\
    RX & 0.10 & 0.06 & \textbf{0.10} & -0.22 \\
    IQP & 0.14 & -0.02 & \textbf{0.10} & -0.34 \\
    QAOA & 0.26 & 0.14 & -0.06 & \textbf{-1.92}\\
    \hline
    \end{tabularx}
\caption{Total difference by metric, over all unique combinations containing each base kernel. Every entry is a sum of six elements from the difference grid for the given metric in final column of Fig. \ref{fig:metrics_a-r-m-s}. Boldface entries indicate the kernel(s) showing the greatest total difference by magnitude, for each metric. Improvements in the accuracy, AUCROC, and margin correspond to positive values (\textit{i.e.} a total increase). However, a total increase \textit{or} decrease can both indicate an improvement in the spectral ratio, depending on initial values in the first column of Fig. \ref{fig:metrics_a-r-m-s}.}
\label{table:metrcs_a-r-m-s_sum_improvement}
\end{table*}

\subsubsection{AUCROC}
Trends in the AUCROC values are broadly comparable to trends in the accuracy, as evident in the first two rows of Fig. \ref{fig:metrics_a-r-m-s}. Here, again, the RBF- and QAOA-containing kernels show the largest overall improvement. Individually, the largest improvements are for the QAOA base kernel and the IQP-RBF kernel combination. A significant \textit{decrease} in the AUCROC is seen here for the IQP-Linear and IQP-Polynomial kernel combinations, despite a net-zero change in the accuracy, when comparing (i) and (iii). Noting the low median AUCROC of the IQP base kernel (0.71), this suggests a stronger preference for the IQP kernel versus both the Linear and Polynomial kernels \textit{vis a vis} the optimization target Eq. (\ref{eq:separation-component}), which is not necessarily aligned with the AUCROC metric. The same line of reasoning applies for the Linear and Polynomial kernels in combination with QAOA. Conversely, a stronger preference RBF in the IQP-RBF combination (in addition to optimizing $\theta_2$ (Tab. \ref{table:classical-kernels})) leads to a large \textit{increase} (+0.08) in the AUCROC over the balanced IQP-RBF combination. It is also noteworthy that, despite a lower score overall, the base QAOA kernel exhibits a comparatively strong improvement in the median AUCROC with parameter optimization.

\subsubsection{Margin}
Outcomes for the margin metric also show the largest differences among combinations that pair low- and high-scoring base kernels. Excluding combinations with the two lowest-scoring base kernels (Linear, Polynomial), a slight overall decrease in the margin is evident among the remaining kernel combinations. Moreover, negative changes are associated with the RBF- and QAOA-containing combinations, and most strongly with the lone base kernels in either case. This suggests that parameter optimization tends to simultaneously reduce the margin, despite generally improving accuracy and AUCROC.

\subsubsection{Spectral Ratio}
We observe the largest differences overall in comparing type (i) and type (iii) results for the spectral ratio. Selection against the Linear and Polynomial kernels during $\bm{\gamma}$ optimization leads to a comparative increase of the spectral ratios here for related combinations, because the matching base kernels exhibit very low spectral ratios in general.
\begin{figure*}[t]
    \centering
    \includegraphics[width=0.85\linewidth]{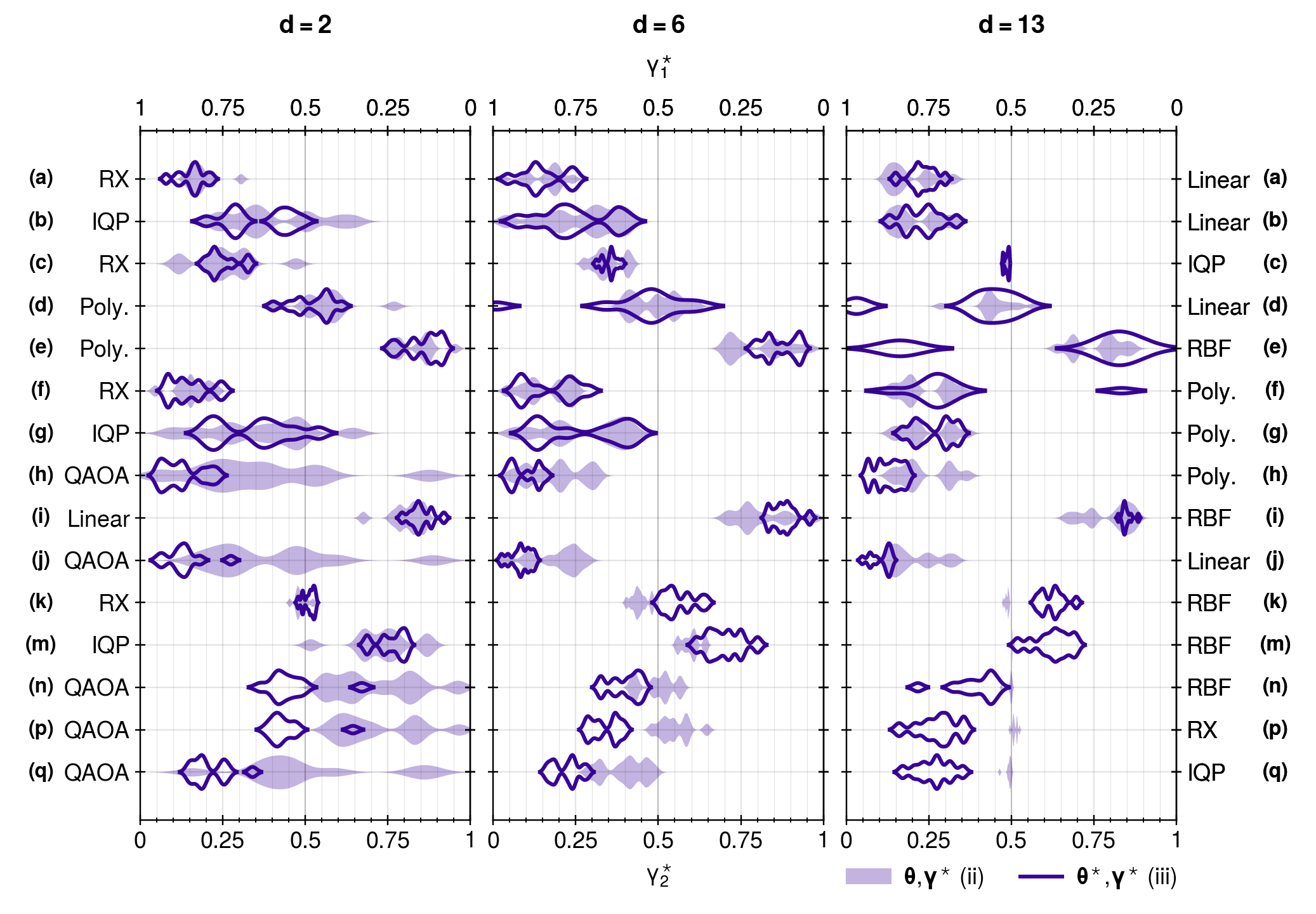}
    \caption{Density of optimized kernel weights ($\gamma_1^\star + \gamma_2^\star = 1$) with and without additional optimization of kernel parameters ($\bm{\theta}$), for $d \in \{2,6,13\}$ features. For a given $d$, distributions skewing right indicate a preference for the kernel on the right-hand side ($\gamma_2^\star > \gamma_1^\star$), and vice versa when skewing left ($\gamma_1^\star > \gamma_2^\star$). Outlined distributions correspond to fully optimized results (type (iii)) and filled distributions correspond to \textit{semi-optimized} results (type (ii)). Alphabetic labels are provided for convenience.}
    \label{fig:kernel-weights-2-6-13}
\end{figure*}
Excluding Linear- and Polynomial-containing entries, all combinations among the RBF, QAOA, RX, and IQP kernels exhibit very high spectral ratios in the ``no optimization'' column, (i). Without tunable parameters, there is little expectation of any change for the RX and IQP base kernels, as well as the RX-IQP combination, because adjusting the kernel weights only interpolates between two already-large values. Indeed, the spectral ratio for RX- and IQP- containing kernels does not change significantly from column (i) to (iii). For combinations that contain the parametric RBF or QAOA kernels, on the other hand, a strong decrease in the spectral ratio toward more moderate values ($\approx 0.5$) is clearly observed. Moreover, the existence and severity of this trend for the \textit{lone} RBF and QAOA kernels (third and final diagonal entries, respectively, in the bottom right grid of Fig. \ref{fig:metrics_a-r-m-s}) confirms that parameter optimization is effective at balancing the spectral ratio.

\subsubsection{Summary}
Overall, these results show that kernel combinations, especially those involving parametric kernels, benefit significantly from the \textit{QCC-net} optimization procedure. The RBF- and RX-containing kernels demonstrated the highest accuracy and AUCROC scores, with RBF- and QAOA-containing kernels exhibiting the largest \textit{improvements} for these metrics. Specifically, the IQP-RBF kernel combination displayed the greatest overall improvement in accuracy, which we attribute to the regularization effect of the smoother RBF kernel. Conversely, the IQP, Linear, and Polynomial kernels showed lower overall performance. The results also highlight that parameter optimization can effectively balance the spectral ratio, which is particularly important for parametric kernels like RBF and QAOA.

\subsection{Impact of Parameter Training on MKL \label{subsec:training-impact}}
In order to separate the effects of kernel parameter training and kernel weights optimization, we proceed with a comparison between type (ii) and type (iii) results (Tab. \ref{table:result-types}). Recall that the latter uses \textit{both} trained $\bm{\theta}^\star$ and optimized $\bm{\gamma}^\star$, whereas the former uses \textit{random or default} $\bm{\theta}$ and optimized $\bm{\gamma}^\star$. We shall outline the comparison here in terms of the optimal kernel weights determined in either case. As illustrated in Fig. \ref{fig:kernel-weights-2-6-13}, we compare the distributions of $\bm{\gamma}^\star$ for data with $d=2$, $6$, and $13$ features to distinguish, also, any trends in $\bm{\gamma}^\star$ that depend on $d$.

\subsubsection{Non-Parametric Combinations}
To start, we confirm that type (ii) and type (iii) results are indeed very similar for kernel combinations that contain no parametric base kernels, namely those on rows \textbf{(a)} to \textbf{(c)}. We can therefore discuss this subset of outcomes without distinguishing between the two types of results. A clear preference for the RX kernel over the Linear kernel is seen at all three values of $d$ on row \textbf{(a)} of Fig. \ref{fig:kernel-weights-2-6-13}. The same can be said for IQP, regarding the IQP-Linear combination on row \textbf{(b)}. Neither result is surprising, since the Linear kernel is obviously not suited to the non-linearly-separable classification problem under consideration (see, for example, Fig. \ref{fig:SM-visualize-2d}). Nonetheless, results from the previous section (Fig. \ref{fig:metrics_a-r-m-s}) indicate that the RX-Linear and IQP-Linear combinations do in fact outperform the lone quantum kernels in either case, which may explain why the Linear kernel is not eliminated entirely (\textit{i.e.} $\gamma_2 \neq 0$). On row \textbf{(c)}, preference for the RX kernel is strong at $d=2$ for this non-parametric q.q. combination. However, a gradual shift toward balanced weights, and a narrowing of the weights distribution, is observed with increasing $d$. For the largest number of features, $d=13$, the RX and IQP kernels appear equally effective.

\subsubsection{Parametric Kernel Combinations}
In contrast to the above, results for the c.c. Polynomial-Linear combination, \textbf{(d)}, show approximately balanced weights across all $d$, apart from a small proportion of the fully optimized outcomes (iii) that skew very strongly toward the Polynomial kernel. This suggests rare instances in which parameter training is highly successful for the Polynomial kernel. Results for the Linear-RBF combination immediately below, on row \textbf{(e)}, may further support this: While the RBF kernel largely dominates the combination, a comparable proportion of the outcomes at the $d=13$ strongly favour the fully optimized Polynomial kernel. This is \textit{not} to suggest, however, that the trained Polynomial kernel is particularly effective for the problem at hand, since its combination with the RX and IQP kernels, on rows \textbf{(f)} and \textbf{(g)}, exhibits minimal difference between result types (ii) and (iii).

The QAOA-Polynomial combination on row \textbf{(h)} corresponds to the first unambiguous result as far as differentiating trained and random-parameter outcomes. Preference for the QAOA kernel over the Linear kernel is clear with and without parameter training, although much narrower distributions are observed for the former (type (iii)), especially at lower $d$. A similar trend is seen for the QAOA-Linear combination on row \textbf{(j)}. With the Linear-RBF combination in between the prior two results, on row \textbf{(i)}, the trend is again similar, except the semi-optimized distributions are more narrow here to start. Evidently, the initial value of the RBF scaling parameter, $\theta_0 = 1$, represents a reasonable choice for data with features scaled to $[0, 2\pi]$. No such choice can be made for the QAOA kernel, however, which is parametrized via periodic quantum gates. Thus, for the type (ii) results on rows \textbf{(h)} and \textbf{(j)}, random-parameter outcomes for QAOA-containing combinations are more broadly distributed compared to the Linear-RBF combination.

\subsubsection{Quantum-RBF and QAOA-Quantum Kernel Combinations}
Rows \textbf{(k)} to \textbf{(q)} correspond to pairs among RBF and the three quantum kernels (Tab. \ref{table:quantum-kernels}). For the RX-RBF combination, \textbf{(k)}, we note that $\bm{\gamma}^\star$ remains approximately balanced across $d$ without parameter optimization. For type (iii) results, however, the RBF kernel is preferred over RX, and more so with increasing $d$. Next, for IQP-RBF kernel combination on row \textbf{(m)}, we note that $\bm{\gamma}^\star$ skews slightly toward RBF for both types of results at the two lower values of $d$. This is more severe for the fully optimized case (iii), in view of the $d=6$ result. At $d=13$, the semi-optimized case (ii) does not distinguish at all between IQP and RBF. (The distribution in Fig. \ref{fig:kernel-weights-2-6-13} row \textbf{(m)} is very narrow and obscured by the vertical grid line at $\gamma_1^\star = \gamma_2^\star = 0.5$.) In the fully optimized case, however, a preference for the RBF kernel is clearly visible. Here, and for the RX-RBF combination, \textbf{(k)}, parameter training enhances the selectivity for weights tuning, especially at higher dimensions.

The final rows of Fig. \ref{fig:kernel-weights-2-6-13}, \textbf{(n)} to \textbf{(q)}, contain the QAOA kernel in combination with RBF, RX, and IQP, respectively. Regarding the semi-optimized results for these three rows, we note that, while the random-parameter QAOA kernel (ii) is selected \textit{against} for data with $d=2$ features, this trend gradually disappears for larger $d$. Indeed, at $d=13$, rows \textbf{(n)} to \textbf{(q)} illustrate virtually no preference between the random $\bm{\theta}$ QAOA kernel and the RBF, RX, or IQP kernels. The opposite trend is observed, however, for the fully optimized results on these combinations. That is, preference for the fully optimized QAOA kernel increases from $d=2$ to $13$. This result, too, supports the concluding claim in the previous paragraph. Operating upon data with $d=13$ features, the weights tuning algorithm (Sec. \ref{EasyMKL}) does not discriminate between component kernels in neither q.q., nor RBF-containing q.c. combinations with the base kernels considered in this work.

\subsubsection{Summary}
These findings reveal that the RX, IQP, and RBF kernels surpass the Linear and Polynomial kernels for the studied classification problems. Furthermore, the fully optimized QAOA kernel is more strongly preferred as data dimensionality grows, suggesting that parameter training serves to enable kernel weights tuning in high-dimensional settings. The results also show limited discrimination between quantum and classical kernels in q.q. and RBF-containing q.c. combinations, suggesting that kernel and optimization choices should be tailored to the problem and dataset specifics.
\begin{figure*}[t]
     \centering
     \begin{subfigure}[b]{0.425\textwidth}
        \centering
        \includegraphics[width=\textwidth]{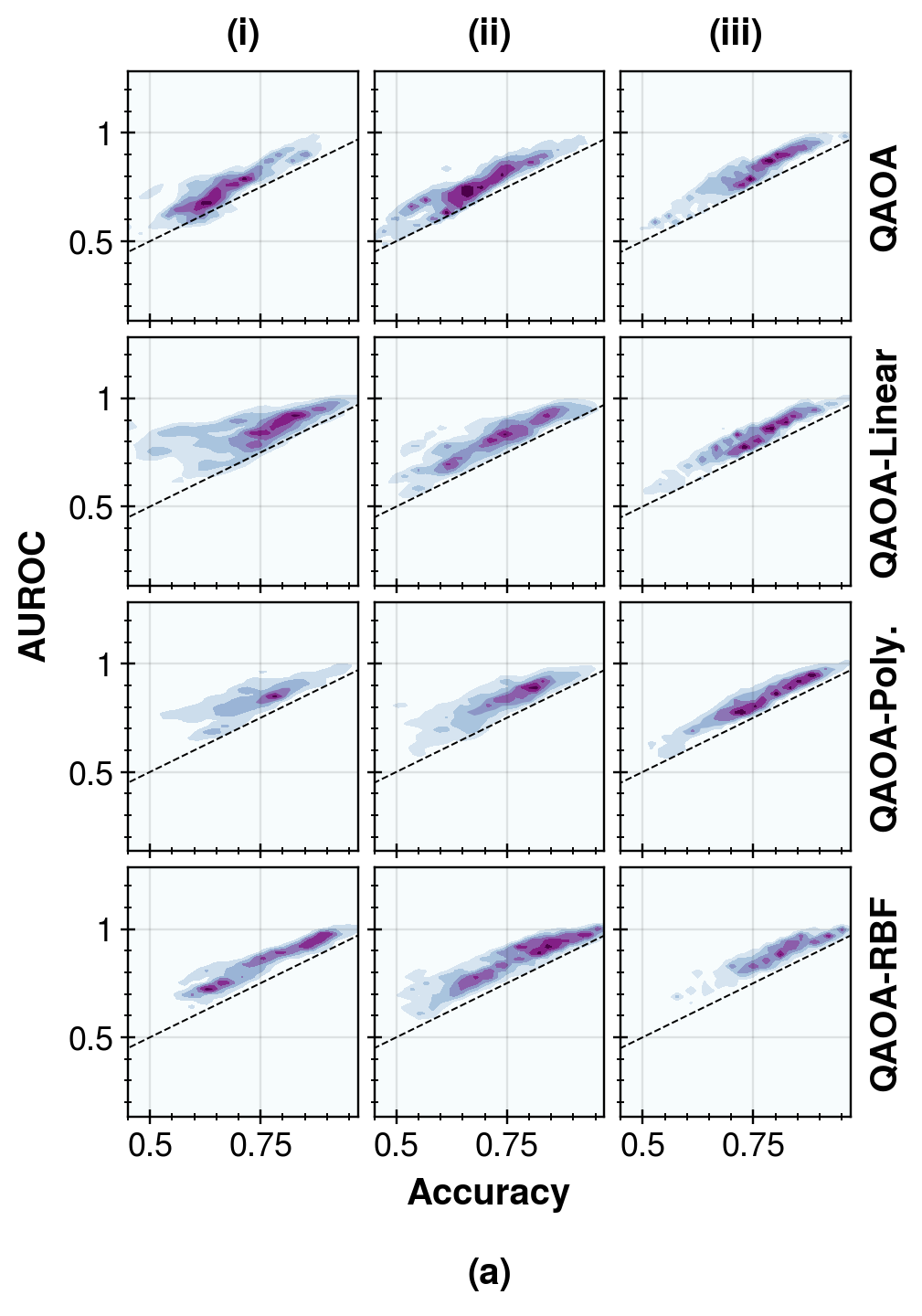}
        \label{fig:density-plot-acc-roc}
     \end{subfigure}
     \hspace{1cm}
     \begin{subfigure}[b]{0.425\textwidth}
        \centering
        \includegraphics[width=\textwidth]{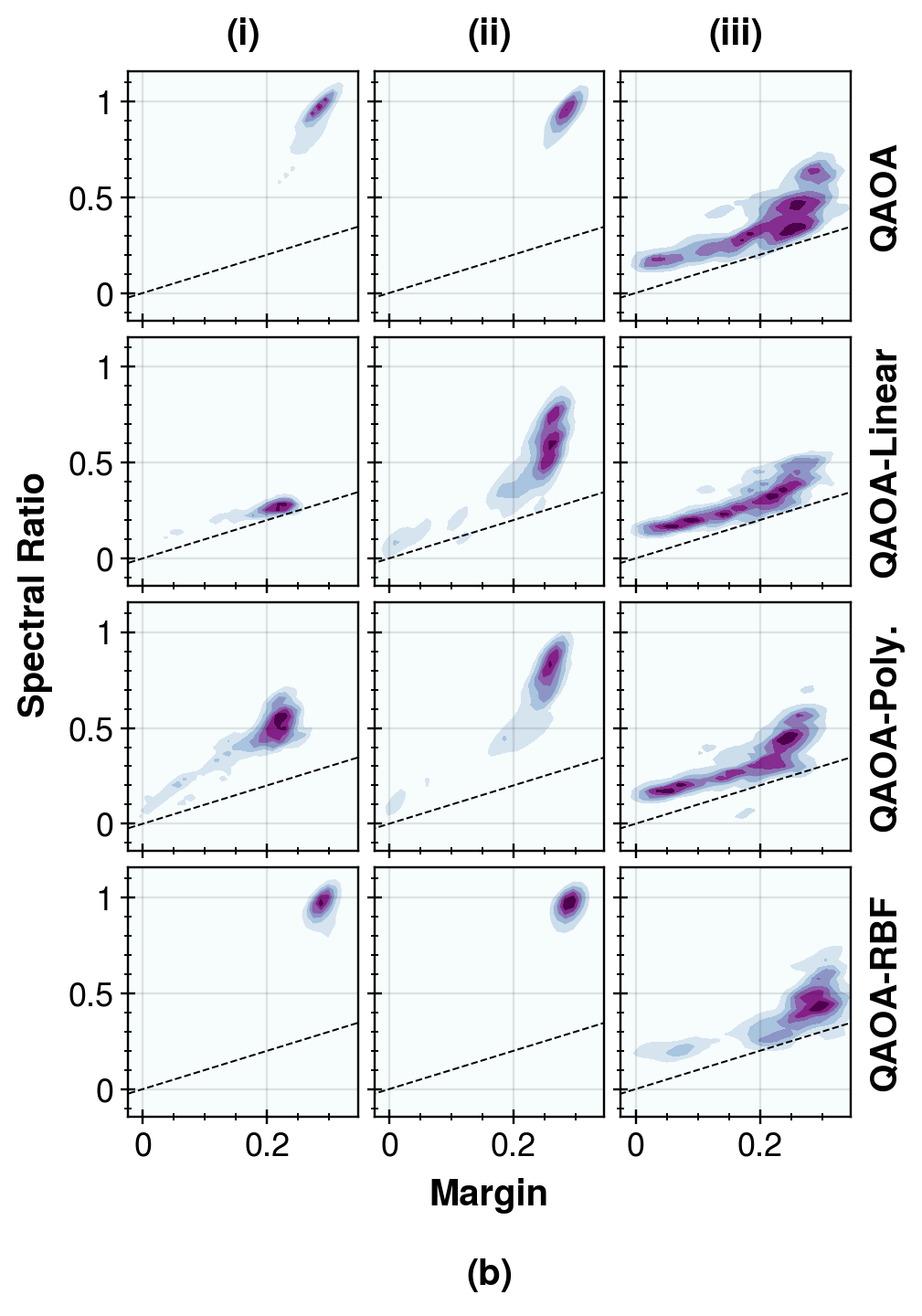}
        \label{fig:density-plot-margin-spectral}
     \end{subfigure}
        \caption{Density contours illustrating distributions of performance metrics for QAOA-containing q.c. combinations for all three result types, using (i) random kernel parameters and equal kernel weights ($\bm{\theta}, \bm{\gamma}$); (ii) random kernel parameters and optimized weights ($\bm{\theta}, \bm{\gamma}^\star$); and (iii) trained kernel parameters and optimized kernel weights ($\bm{\theta}^\star, \bm{\gamma}^\star$). The dashed black lines in all subplots indicate $x=y$ and are provided to help orient to eye. Distributions include results for all feature sizes, $d=2$ to $13$. Note that matching results are expected for cases (i) and (ii) along the top row of either set of subplots, since these correspond to the lone QAOA base kernel.}
        \label{fig:density-plots}
\end{figure*}
\subsection{Insights on QC Combinations}
Generalizing upon results from the previous two sections, the RBF-containing q.c. combinations exhibit the best performance overall and the greatest improvements in performance metrics between result types (i) and (iii). When paired with the quantum kernels from Tab. \ref{table:quantum-kernels}, MKL optimization weighs the untrained RBF kernel equally or more heavily than all three quantum counterparts, including the untrained (random parameter) QAOA kernel. Based on the median metrics in Fig. \ref{fig:metrics_a-r-m-s}, we can confirm that training $\theta_2$ improves the lone RBF kernel's performance. The fact that the \textit{trained} QAOA kernel is weighted more heavily than the trained RBF kernel (Fig. \ref{fig:kernel-weights-2-6-13}, row \textbf{(n)}) then suggests an important and competitive contribution from QAOA. However, even in considering only $d=13$ feature data, median metrics for case (iii) closely resemble those of results across all $d$ in Fig. \ref{fig:metrics_a-r-m-s}, where the lone RBF kernel consistently out-performs the QAOA-RBF combination.

One reason for this could be the shallow depth of the minimal ans\"atze utilized for QAOA and the other quantum kernels (Tab. \ref{table:quantum-kernels}). In the context of its original application \cite{Farhi2014}, multiple repetitions of the QAOA \textit{circuit} are known to produce higher quality approximations to solutions for combinatorial problems. It may be the case that a similar relationship exists for QEKs that also utilize multiple, parametric QAOA layers for classification with kernel methods \cite{jerbi2023quantum}. The related, though non-parametric, IQP kernel could also benefit from repetitions of the embedding circuit---such repetitions are, in fact, a requirement for rigorous classical ``hardness'' in this case \cite{Havlivcek2019}.

The observation of better performance metrics for the RX-containing q.c. combinations, in comparison to the QAOA- and IQP-containing combinations, is not necessarily contradictory to the above hypothesis. Considering that the RX kernel is effectively a quantum implementation of a classical cosine kernel \cite{Schuld2018}, and acknowledging the relative simplicity of the synthetic datasets under consideration, the utility of more complex QEKs may yet be demonstrable on higher-dimensional or specifically structured data, as in Refs. \cite{Liu2021,glick2021covariant}. 

In the present context, QAOA remains the best representative of a trainable and practical family of quantum kernels. The distributions of performance metrics for q.c. combinations that contain this kernel, and those of the base QAOA kernel, are illustrated in Fig. \ref{fig:density-plots}. Here, as also revealed by the median metrics in Fig. \ref{fig:metrics_a-r-m-s}, the overall accuracy and AUCROC are only slightly improved, and the variance of outcomes only slightly reduced, in comparing result types (i) and (iii). Still more subtle is the accuracy and AUCROC difference between types (i) and (ii). The latter is unsurprising in view of Fig. \ref{fig:kernel-weights-2-6-13}, since without parameter training the optimized weights are seen to converge to the balanced (default) vector, $\bm{\gamma}^\star \rightarrow \bm{\gamma} = [0.5, 0.5]^T$, as $d$ increases.

Regarding the margin and spectral ratio outcomes for this subset of q.c. combinations (Fig. \ref{fig:density-plots} \textbf{(b)}), the type (ii) results suggest that weights optimization \textit{without} parameter training may actually be detrimental for the QAOA-Linear and QAOA-Polynomial combinations, whereas the QAOA-RBF distribution is again not significantly altered. With the inclusion of parameter training in case (iii), the distributions for the QAOA kernel and all QAOA-containing q.c. combinations are seen to change more drastically. Specifically, the margin distribution is widened while the spectral ratios are reduced for type (iii) results across every row in Fig. \ref{fig:density-plots} \textbf{(b)}. The type (iii) results here also converge to similar distributions for all four combinations considered, with the QAOA-RBF combination producing the best outcome overall.

\section{Conclusion \label{sec:conclusion}}
In this work, we used modern software tools and a novel optimization procedure inspired by classical machine learning to systematically explore the utility of c.c., q.c., and q.q. kernel combinations in binary classification problems, over datasets with $d=2$ to 13 features.  Considering three quantum kernels and three canonical classical kernels in a comparative setting, we found that only the most complex and parametric quantum kernel (QAOA) attains higher a optimum weight in pairwise combination with the most performant classical kernel (RBF). Conversely, classification performance was not found to differ significantly between q.c. combinations featuring \textit{simpler} classical kernels (Linear, Polynomial) in comparison to the lone quantum kernel. Regarding use of the \textit{QCC-net} for training the kernel parameters and optimizing the combination weights, we found the kernel weighting step to be indecisive without the former training step when optimizing q.c. combinations that contain the parametric QAOA kernel. The simpler RX kernel and its q.c. combinations exhibited the best performance metrics among the quantum and quantum-containing kernels considered. More broadly, classical RBF kernel and its combinations performed the best overall.

We are able to recommend a number of directions for future work aimed at expanding the scope of this study and/or identifying an empirical advantage with quantum kernels. For example, the use of \textit{multi}-layer embedding circuits for quantum kernels may prove more effective on higher-dimensional data, based on the trends we observed for QAOA-containing kernel combinations. Additionally, the \textit{EasyMKL} algorithm is well suited for combining a far greater number of kernels and can therefore be used to explore the q.q, c.c., and q.c paradigms beyond pairwise combinations. On the other hand, alternate (non-linear) MKL strategies are also a worthwhile prospect for future work, assuming that combination weights can be computed efficiently. Finally, experiments on datasets with a different, less generic structure may provide the more promising results for q.q. and q.c. combinations, especially if the number of features is large.\\

\section{Acknowledgements}
Partial funding for this work was provided by the Mitacs Accelerate program.

\bibliography{quantum-classical_mkl.bib}

\begin{thebibliography}{58}%
\makeatletter
\providecommand \@ifxundefined [1]{%
 \@ifx{#1\undefined}
}%
\providecommand \@ifnum [1]{%
 \ifnum #1\expandafter \@firstoftwo
 \else \expandafter \@secondoftwo
 \fi
}%
\providecommand \@ifx [1]{%
 \ifx #1\expandafter \@firstoftwo
 \else \expandafter \@secondoftwo
 \fi
}%
\providecommand \natexlab [1]{#1}%
\providecommand \enquote  [1]{``#1''}%
\providecommand \bibnamefont  [1]{#1}%
\providecommand \bibfnamefont [1]{#1}%
\providecommand \citenamefont [1]{#1}%
\providecommand \href@noop [0]{\@secondoftwo}%
\providecommand \href [0]{\begingroup \@sanitize@url \@href}%
\providecommand \@href[1]{\@@startlink{#1}\@@href}%
\providecommand \@@href[1]{\endgroup#1\@@endlink}%
\providecommand \@sanitize@url [0]{\catcode `\\12\catcode `\$12\catcode
  `\&12\catcode `\#12\catcode `\^12\catcode `\_12\catcode `\%12\relax}%
\providecommand \@@startlink[1]{}%
\providecommand \@@endlink[0]{}%
\providecommand \url  [0]{\begingroup\@sanitize@url \@url }%
\providecommand \@url [1]{\endgroup\@href {#1}{\urlprefix }}%
\providecommand \urlprefix  [0]{URL }%
\providecommand \Eprint [0]{\href }%
\providecommand \doibase [0]{http://dx.doi.org/}%
\providecommand \selectlanguage [0]{\@gobble}%
\providecommand \bibinfo  [0]{\@secondoftwo}%
\providecommand \bibfield  [0]{\@secondoftwo}%
\providecommand \translation [1]{[#1]}%
\providecommand \BibitemOpen [0]{}%
\providecommand \bibitemStop [0]{}%
\providecommand \bibitemNoStop [0]{.\EOS\space}%
\providecommand \EOS [0]{\spacefactor3000\relax}%
\providecommand \BibitemShut  [1]{\csname bibitem#1\endcsname}%
\let\auto@bib@innerbib\@empty
\bibitem [{\citenamefont {Preskill}(2018)}]{Preskill2018}%
  \BibitemOpen
  \bibfield  {author} {\bibinfo {author} {\bibfnamefont {J.}~\bibnamefont
  {Preskill}},\ }\href {\doibase https://doi.org/10.22331/q-2018-08-06-79}
  {\bibfield  {journal} {\bibinfo  {journal} {arXiv preprint arXiv:1801.00862}\
  }\textbf {\bibinfo {volume} {2}},\ \bibinfo {pages} {79} (\bibinfo {year}
  {2018})}\BibitemShut {NoStop}%
\bibitem [{\citenamefont {Arute}\ \emph {et~al.}(2019)\citenamefont {Arute},
  \citenamefont {Arya}, \citenamefont {Babbush}, \citenamefont {Bacon},
  \citenamefont {Bardin}, \citenamefont {Barends}, \citenamefont {Biswas},
  \citenamefont {Boixo}, \citenamefont {Brandao}, \citenamefont {Buell} \emph
  {et~al.}}]{arute2019quantum}%
  \BibitemOpen
  \bibfield  {author} {\bibinfo {author} {\bibfnamefont {F.}~\bibnamefont
  {Arute}}, \bibinfo {author} {\bibfnamefont {K.}~\bibnamefont {Arya}},
  \bibinfo {author} {\bibfnamefont {R.}~\bibnamefont {Babbush}}, \bibinfo
  {author} {\bibfnamefont {D.}~\bibnamefont {Bacon}}, \bibinfo {author}
  {\bibfnamefont {J.~C.}\ \bibnamefont {Bardin}}, \bibinfo {author}
  {\bibfnamefont {R.}~\bibnamefont {Barends}}, \bibinfo {author} {\bibfnamefont
  {R.}~\bibnamefont {Biswas}}, \bibinfo {author} {\bibfnamefont
  {S.}~\bibnamefont {Boixo}}, \bibinfo {author} {\bibfnamefont {F.~G.}\
  \bibnamefont {Brandao}}, \bibinfo {author} {\bibfnamefont {D.~A.}\
  \bibnamefont {Buell}},  \emph {et~al.},\ }\href {\doibase
  https://doi.org/10.1038/s41586-019-1666-5} {\bibfield  {journal} {\bibinfo
  {journal} {Nature}\ }\textbf {\bibinfo {volume} {574}},\ \bibinfo {pages}
  {505} (\bibinfo {year} {2019})}\BibitemShut {NoStop}%
\bibitem [{\citenamefont {Madsen}\ \emph {et~al.}(2022)\citenamefont {Madsen},
  \citenamefont {Laudenbach}, \citenamefont {Askarani}, \citenamefont
  {Rortais}, \citenamefont {Vincent}, \citenamefont {Bulmer}, \citenamefont
  {Miatto}, \citenamefont {Neuhaus}, \citenamefont {Helt}, \citenamefont
  {Collins} \emph {et~al.}}]{madsen2022quantum}%
  \BibitemOpen
  \bibfield  {author} {\bibinfo {author} {\bibfnamefont {L.~S.}\ \bibnamefont
  {Madsen}}, \bibinfo {author} {\bibfnamefont {F.}~\bibnamefont {Laudenbach}},
  \bibinfo {author} {\bibfnamefont {M.~F.}\ \bibnamefont {Askarani}}, \bibinfo
  {author} {\bibfnamefont {F.}~\bibnamefont {Rortais}}, \bibinfo {author}
  {\bibfnamefont {T.}~\bibnamefont {Vincent}}, \bibinfo {author} {\bibfnamefont
  {J.~F.}\ \bibnamefont {Bulmer}}, \bibinfo {author} {\bibfnamefont {F.~M.}\
  \bibnamefont {Miatto}}, \bibinfo {author} {\bibfnamefont {L.}~\bibnamefont
  {Neuhaus}}, \bibinfo {author} {\bibfnamefont {L.~G.}\ \bibnamefont {Helt}},
  \bibinfo {author} {\bibfnamefont {M.~J.}\ \bibnamefont {Collins}},  \emph
  {et~al.},\ }\href {\doibase https://doi.org/10.1038/s41586-022-04725-x}
  {\bibfield  {journal} {\bibinfo  {journal} {Nature}\ }\textbf {\bibinfo
  {volume} {606}},\ \bibinfo {pages} {75} (\bibinfo {year} {2022})}\BibitemShut
  {NoStop}%
\bibitem [{\citenamefont {Shawe-Taylor}\ and\ \citenamefont
  {Cristianini}(2004)}]{ShaweTaylor2004}%
  \BibitemOpen
  \bibfield  {author} {\bibinfo {author} {\bibfnamefont {J.}~\bibnamefont
  {Shawe-Taylor}}\ and\ \bibinfo {author} {\bibfnamefont {N.}~\bibnamefont
  {Cristianini}},\ }\href {\doibase https://doi.org/10.1017/cbo9780511809682}
  {\emph {\bibinfo {title} {Kernel Methods for Pattern Analysis}}}\ (\bibinfo
  {publisher} {Cambridge University Press},\ \bibinfo {year}
  {2004})\BibitemShut {NoStop}%
\bibitem [{\citenamefont {Biamonte}\ \emph {et~al.}(2017)\citenamefont
  {Biamonte}, \citenamefont {Wittek}, \citenamefont {Pancotti}, \citenamefont
  {Rebentrost}, \citenamefont {Wiebe},\ and\ \citenamefont
  {Lloyd}}]{Biamonte2017}%
  \BibitemOpen
  \bibfield  {author} {\bibinfo {author} {\bibfnamefont {J.}~\bibnamefont
  {Biamonte}}, \bibinfo {author} {\bibfnamefont {P.}~\bibnamefont {Wittek}},
  \bibinfo {author} {\bibfnamefont {N.}~\bibnamefont {Pancotti}}, \bibinfo
  {author} {\bibfnamefont {P.}~\bibnamefont {Rebentrost}}, \bibinfo {author}
  {\bibfnamefont {N.}~\bibnamefont {Wiebe}}, \ and\ \bibinfo {author}
  {\bibfnamefont {S.}~\bibnamefont {Lloyd}},\ }\href {\doibase
  https://doi.org/10.1038/nature23474} {\bibfield  {journal} {\bibinfo
  {journal} {Nature}\ }\textbf {\bibinfo {volume} {549}},\ \bibinfo {pages}
  {195} (\bibinfo {year} {2017})}\BibitemShut {NoStop}%
\bibitem [{\citenamefont {Cortes}\ and\ \citenamefont
  {Vapnik}(1995)}]{cortes1995support}%
  \BibitemOpen
  \bibfield  {author} {\bibinfo {author} {\bibfnamefont {C.}~\bibnamefont
  {Cortes}}\ and\ \bibinfo {author} {\bibfnamefont {V.}~\bibnamefont
  {Vapnik}},\ }\href {\doibase https://doi.org/10.1007/BF00994018} {\bibfield
  {journal} {\bibinfo  {journal} {Mach. Learn.}\ }\textbf {\bibinfo {volume}
  {20}},\ \bibinfo {pages} {273} (\bibinfo {year} {1995})}\BibitemShut
  {NoStop}%
\bibitem [{\citenamefont {Mengoni}\ and\ \citenamefont
  {Pierro}(2019)}]{Mengoni2019}%
  \BibitemOpen
  \bibfield  {author} {\bibinfo {author} {\bibfnamefont {R.}~\bibnamefont
  {Mengoni}}\ and\ \bibinfo {author} {\bibfnamefont {A.~D.}\ \bibnamefont
  {Pierro}},\ }\href {\doibase https://doi.org/10.1007/s42484-019-00007-4}
  {\bibfield  {journal} {\bibinfo  {journal} {Quantum Mach. Intell.}\ }\textbf
  {\bibinfo {volume} {1}},\ \bibinfo {pages} {65} (\bibinfo {year}
  {2019})}\BibitemShut {NoStop}%
\bibitem [{\citenamefont {Rebentrost}\ \emph {et~al.}(2014)\citenamefont
  {Rebentrost}, \citenamefont {Mohseni},\ and\ \citenamefont
  {Lloyd}}]{rebentrost2014quantum}%
  \BibitemOpen
  \bibfield  {author} {\bibinfo {author} {\bibfnamefont {P.}~\bibnamefont
  {Rebentrost}}, \bibinfo {author} {\bibfnamefont {M.}~\bibnamefont {Mohseni}},
  \ and\ \bibinfo {author} {\bibfnamefont {S.}~\bibnamefont {Lloyd}},\ }\href
  {\doibase https://doi.org/10.1103/PhysRevLett.113.130503} {\bibfield
  {journal} {\bibinfo  {journal} {Phys. Rev. Lett.}\ }\textbf {\bibinfo
  {volume} {113}},\ \bibinfo {pages} {130503} (\bibinfo {year}
  {2014})}\BibitemShut {NoStop}%
\bibitem [{\citenamefont {Liu}\ \emph {et~al.}(2021)\citenamefont {Liu},
  \citenamefont {Arunachalam},\ and\ \citenamefont {Temme}}]{Liu2021}%
  \BibitemOpen
  \bibfield  {author} {\bibinfo {author} {\bibfnamefont {Y.}~\bibnamefont
  {Liu}}, \bibinfo {author} {\bibfnamefont {S.}~\bibnamefont {Arunachalam}}, \
  and\ \bibinfo {author} {\bibfnamefont {K.}~\bibnamefont {Temme}},\ }\href
  {\doibase https://doi.org/10.1038/s41567-021-01287-z} {\bibfield  {journal}
  {\bibinfo  {journal} {Nat. Phys.}\ }\textbf {\bibinfo {volume} {17}},\
  \bibinfo {pages} {1013} (\bibinfo {year} {2021})}\BibitemShut {NoStop}%
\bibitem [{\citenamefont {Peters}\ \emph {et~al.}(2021)\citenamefont {Peters},
  \citenamefont {Caldeira}, \citenamefont {Ho}, \citenamefont {Leichenauer},
  \citenamefont {Mohseni}, \citenamefont {Neven}, \citenamefont {Spentzouris},
  \citenamefont {Strain},\ and\ \citenamefont {Perdue}}]{peters2021machine}%
  \BibitemOpen
  \bibfield  {author} {\bibinfo {author} {\bibfnamefont {E.}~\bibnamefont
  {Peters}}, \bibinfo {author} {\bibfnamefont {J.}~\bibnamefont {Caldeira}},
  \bibinfo {author} {\bibfnamefont {A.}~\bibnamefont {Ho}}, \bibinfo {author}
  {\bibfnamefont {S.}~\bibnamefont {Leichenauer}}, \bibinfo {author}
  {\bibfnamefont {M.}~\bibnamefont {Mohseni}}, \bibinfo {author} {\bibfnamefont
  {H.}~\bibnamefont {Neven}}, \bibinfo {author} {\bibfnamefont
  {P.}~\bibnamefont {Spentzouris}}, \bibinfo {author} {\bibfnamefont
  {D.}~\bibnamefont {Strain}}, \ and\ \bibinfo {author} {\bibfnamefont {G.~N.}\
  \bibnamefont {Perdue}},\ }\href {\doibase
  https://doi.org/10.1038/s41534-021-00498-9} {\bibfield  {journal} {\bibinfo
  {journal} {Npj Quantum Inf.}\ }\textbf {\bibinfo {volume} {7}},\ \bibinfo
  {pages} {161} (\bibinfo {year} {2021})}\BibitemShut {NoStop}%
\bibitem [{\citenamefont {Sancho-Lorente}\ \emph {et~al.}(2022)\citenamefont
  {Sancho-Lorente}, \citenamefont {Rom{\'a}n-Roche},\ and\ \citenamefont
  {Zueco}}]{sancho2022quantum}%
  \BibitemOpen
  \bibfield  {author} {\bibinfo {author} {\bibfnamefont {T.}~\bibnamefont
  {Sancho-Lorente}}, \bibinfo {author} {\bibfnamefont {J.}~\bibnamefont
  {Rom{\'a}n-Roche}}, \ and\ \bibinfo {author} {\bibfnamefont {D.}~\bibnamefont
  {Zueco}},\ }\href {\doibase https://doi.org/10.1103/PhysRevA.105.042432}
  {\bibfield  {journal} {\bibinfo  {journal} {Phys. Rev. A}\ }\textbf {\bibinfo
  {volume} {105}},\ \bibinfo {pages} {042432} (\bibinfo {year}
  {2022})}\BibitemShut {NoStop}%
\bibitem [{\citenamefont {Kyriienko}\ and\ \citenamefont
  {Magnusson}(2022)}]{kyriienko2022unsupervised}%
  \BibitemOpen
  \bibfield  {author} {\bibinfo {author} {\bibfnamefont {O.}~\bibnamefont
  {Kyriienko}}\ and\ \bibinfo {author} {\bibfnamefont {E.~B.}\ \bibnamefont
  {Magnusson}},\ }\href {\doibase https://doi.org/10.48550/arXiv.2208.01203}
  {\bibfield  {journal} {\bibinfo  {journal} {arXiv preprint arXiv:2208.01203}\
  } (\bibinfo {year} {2022}),\
  https://doi.org/10.48550/arXiv.2208.01203}\BibitemShut {NoStop}%
\bibitem [{\citenamefont {Radha}\ and\ \citenamefont
  {Jao}(2022)}]{radha2022generalized}%
  \BibitemOpen
  \bibfield  {author} {\bibinfo {author} {\bibfnamefont {S.~K.}\ \bibnamefont
  {Radha}}\ and\ \bibinfo {author} {\bibfnamefont {C.}~\bibnamefont {Jao}},\
  }\href {\doibase https://doi.org/10.48550/arXiv.2201.02310} {\bibfield
  {journal} {\bibinfo  {journal} {arXiv preprint arXiv:2201.02310}\ } (\bibinfo
  {year} {2022}),\ https://doi.org/10.48550/arXiv.2201.02310}\BibitemShut
  {NoStop}%
\bibitem [{\citenamefont {Schuld}\ and\ \citenamefont
  {Killoran}(2019)}]{Schuld2019}%
  \BibitemOpen
  \bibfield  {author} {\bibinfo {author} {\bibfnamefont {M.}~\bibnamefont
  {Schuld}}\ and\ \bibinfo {author} {\bibfnamefont {N.}~\bibnamefont
  {Killoran}},\ }\href {\doibase
  https://doi.org/10.1103/PhysRevLett.122.040504} {\bibfield  {journal}
  {\bibinfo  {journal} {Phys. Rev. Lett.}\ }\textbf {\bibinfo {volume} {122}},\
  \bibinfo {pages} {040504} (\bibinfo {year} {2019})}\BibitemShut {NoStop}%
\bibitem [{\citenamefont {Hubregtsen}\ \emph {et~al.}(2022)\citenamefont
  {Hubregtsen}, \citenamefont {Wierichs}, \citenamefont {Gil-Fuster},
  \citenamefont {Derks}, \citenamefont {Faehrmann},\ and\ \citenamefont
  {Meyer}}]{Hubregtsen2022}%
  \BibitemOpen
  \bibfield  {author} {\bibinfo {author} {\bibfnamefont {T.}~\bibnamefont
  {Hubregtsen}}, \bibinfo {author} {\bibfnamefont {D.}~\bibnamefont
  {Wierichs}}, \bibinfo {author} {\bibfnamefont {E.}~\bibnamefont
  {Gil-Fuster}}, \bibinfo {author} {\bibfnamefont {P.-J. H.~S.}\ \bibnamefont
  {Derks}}, \bibinfo {author} {\bibfnamefont {P.~K.}\ \bibnamefont
  {Faehrmann}}, \ and\ \bibinfo {author} {\bibfnamefont {J.~J.}\ \bibnamefont
  {Meyer}},\ }\href {\doibase https://doi.org/10.1103/PhysRevA.106.042431}
  {\bibfield  {journal} {\bibinfo  {journal} {Phys. Rev. A}\ }\textbf {\bibinfo
  {volume} {106}},\ \bibinfo {pages} {042431} (\bibinfo {year}
  {2022})}\BibitemShut {NoStop}%
\bibitem [{\citenamefont {G{\"o}nen}\ and\ \citenamefont
  {Alpayd{\i}n}(2011)}]{Gonen2011}%
  \BibitemOpen
  \bibfield  {author} {\bibinfo {author} {\bibfnamefont {M.}~\bibnamefont
  {G{\"o}nen}}\ and\ \bibinfo {author} {\bibfnamefont {E.}~\bibnamefont
  {Alpayd{\i}n}},\ }\href
  {https://www.jmlr.org/papers/volume12/gonen11a/gonen11a.pdf} {\bibfield
  {journal} {\bibinfo  {journal} {J. Mach. Learn. Res.}\ }\textbf {\bibinfo
  {volume} {12}},\ \bibinfo {pages} {2211} (\bibinfo {year}
  {2011})}\BibitemShut {NoStop}%
\bibitem [{\citenamefont {Aiolli}\ and\ \citenamefont
  {Donini}(2015)}]{Aiolli2015}%
  \BibitemOpen
  \bibfield  {author} {\bibinfo {author} {\bibfnamefont {F.}~\bibnamefont
  {Aiolli}}\ and\ \bibinfo {author} {\bibfnamefont {M.}~\bibnamefont
  {Donini}},\ }\href {\doibase https://doi.org/10.1016/j.neucom.2014.11.078}
  {\bibfield  {journal} {\bibinfo  {journal} {Neurocomputing}\ }\textbf
  {\bibinfo {volume} {169}},\ \bibinfo {pages} {215} (\bibinfo {year}
  {2015})}\BibitemShut {NoStop}%
\bibitem [{\citenamefont {Rakotomamonjy}\ \emph {et~al.}(2008)\citenamefont
  {Rakotomamonjy}, \citenamefont {Bach}, \citenamefont {Canu},\ and\
  \citenamefont {Grandvalet}}]{rakotomamonjy2008simplemkl}%
  \BibitemOpen
  \bibfield  {author} {\bibinfo {author} {\bibfnamefont {A.}~\bibnamefont
  {Rakotomamonjy}}, \bibinfo {author} {\bibfnamefont {F.}~\bibnamefont {Bach}},
  \bibinfo {author} {\bibfnamefont {S.}~\bibnamefont {Canu}}, \ and\ \bibinfo
  {author} {\bibfnamefont {Y.}~\bibnamefont {Grandvalet}},\ }\href
  {https://www.jmlr.org/papers/v9/rakotomamonjy08a.html} {\bibfield  {journal}
  {\bibinfo  {journal} {J. Mach. Learn. Res.}\ }\textbf {\bibinfo {volume}
  {9}},\ \bibinfo {pages} {2491} (\bibinfo {year} {2008})}\BibitemShut
  {NoStop}%
\bibitem [{\citenamefont {Suzuki}\ and\ \citenamefont
  {Sugiyama}(2012)}]{suzuki2012fast}%
  \BibitemOpen
  \bibfield  {author} {\bibinfo {author} {\bibfnamefont {T.}~\bibnamefont
  {Suzuki}}\ and\ \bibinfo {author} {\bibfnamefont {M.}~\bibnamefont
  {Sugiyama}},\ }in\ \href {\doibase https://doi.org/10.1214/13-AOS1095} {\emph
  {\bibinfo {booktitle} {Artificial Intelligence and Statistics}}}\ (\bibinfo
  {organization} {PMLR},\ \bibinfo {year} {2012})\ pp.\ \bibinfo {pages}
  {1152--1183}\BibitemShut {NoStop}%
\bibitem [{\citenamefont {Bach}\ \emph {et~al.}(2004)\citenamefont {Bach},
  \citenamefont {Lanckriet},\ and\ \citenamefont {Jordan}}]{bach2004multiple}%
  \BibitemOpen
  \bibfield  {author} {\bibinfo {author} {\bibfnamefont {F.~R.}\ \bibnamefont
  {Bach}}, \bibinfo {author} {\bibfnamefont {G.~R.}\ \bibnamefont {Lanckriet}},
  \ and\ \bibinfo {author} {\bibfnamefont {M.~I.}\ \bibnamefont {Jordan}},\
  }in\ \href {\doibase https://doi.org/10.1145/1015330.1015424} {\emph
  {\bibinfo {booktitle} {Proceedings of the twenty-first international
  conference on Machine learning}}}\ (\bibinfo {year} {2004})\ p.~\bibinfo
  {pages} {6}\BibitemShut {NoStop}%
\bibitem [{\citenamefont {Yang}\ \emph {et~al.}(2011)\citenamefont {Yang},
  \citenamefont {Tang}, \citenamefont {Zhang}, \citenamefont {Lin},
  \citenamefont {Li},\ and\ \citenamefont {Yang}}]{yang2011multiple}%
  \BibitemOpen
  \bibfield  {author} {\bibinfo {author} {\bibfnamefont {Z.}~\bibnamefont
  {Yang}}, \bibinfo {author} {\bibfnamefont {N.}~\bibnamefont {Tang}}, \bibinfo
  {author} {\bibfnamefont {X.}~\bibnamefont {Zhang}}, \bibinfo {author}
  {\bibfnamefont {H.}~\bibnamefont {Lin}}, \bibinfo {author} {\bibfnamefont
  {Y.}~\bibnamefont {Li}}, \ and\ \bibinfo {author} {\bibfnamefont
  {Z.}~\bibnamefont {Yang}},\ }\href {\doibase
  https://doi.org/10.1016/j.artmed.2010.12.002} {\bibfield  {journal} {\bibinfo
   {journal} {Artif. Intell. Med.}\ }\textbf {\bibinfo {volume} {51}},\
  \bibinfo {pages} {163} (\bibinfo {year} {2011})}\BibitemShut {NoStop}%
\bibitem [{\citenamefont {Bach}\ \emph {et~al.}(2012)\citenamefont {Bach},
  \citenamefont {Jenatton}, \citenamefont {Mairal}, \citenamefont {Obozinski}
  \emph {et~al.}}]{bach2012optimization}%
  \BibitemOpen
  \bibfield  {author} {\bibinfo {author} {\bibfnamefont {F.}~\bibnamefont
  {Bach}}, \bibinfo {author} {\bibfnamefont {R.}~\bibnamefont {Jenatton}},
  \bibinfo {author} {\bibfnamefont {J.}~\bibnamefont {Mairal}}, \bibinfo
  {author} {\bibfnamefont {G.}~\bibnamefont {Obozinski}},  \emph {et~al.},\
  }\href {\doibase https://doi.org/10.1561/2200000015} {\bibfield  {journal}
  {\bibinfo  {journal} {Found. Trends Mach. Learn.}\ }\textbf {\bibinfo
  {volume} {4}},\ \bibinfo {pages} {1} (\bibinfo {year} {2012})}\BibitemShut
  {NoStop}%
\bibitem [{\citenamefont {Willsch}\ \emph {et~al.}(2020)\citenamefont
  {Willsch}, \citenamefont {Willsch}, \citenamefont {De~Raedt},\ and\
  \citenamefont {Michielsen}}]{willsch2020support}%
  \BibitemOpen
  \bibfield  {author} {\bibinfo {author} {\bibfnamefont {D.}~\bibnamefont
  {Willsch}}, \bibinfo {author} {\bibfnamefont {M.}~\bibnamefont {Willsch}},
  \bibinfo {author} {\bibfnamefont {H.}~\bibnamefont {De~Raedt}}, \ and\
  \bibinfo {author} {\bibfnamefont {K.}~\bibnamefont {Michielsen}},\ }\href
  {\doibase https://doi.org/10.1016/j.cpc.2019.107006} {\bibfield  {journal}
  {\bibinfo  {journal} {Comput. Phys. Commun.}\ }\textbf {\bibinfo {volume}
  {248}},\ \bibinfo {pages} {107006} (\bibinfo {year} {2020})}\BibitemShut
  {NoStop}%
\bibitem [{\citenamefont {Vedaie}\ \emph {et~al.}(2020)\citenamefont {Vedaie},
  \citenamefont {Noori}, \citenamefont {Oberoi}, \citenamefont {Sanders},\ and\
  \citenamefont {Zahedinejad}}]{Vedaie2020}%
  \BibitemOpen
  \bibfield  {author} {\bibinfo {author} {\bibfnamefont {S.~S.}\ \bibnamefont
  {Vedaie}}, \bibinfo {author} {\bibfnamefont {M.}~\bibnamefont {Noori}},
  \bibinfo {author} {\bibfnamefont {J.~S.}\ \bibnamefont {Oberoi}}, \bibinfo
  {author} {\bibfnamefont {B.~C.}\ \bibnamefont {Sanders}}, \ and\ \bibinfo
  {author} {\bibfnamefont {E.}~\bibnamefont {Zahedinejad}},\ }\href {\doibase
  https://doi.org/10.48550/ARXIV.2011.09694} {\bibfield  {journal} {\bibinfo
  {journal} {arXiv preprint arXiv:2011.09694}\ } (\bibinfo {year} {2020}),\
  https://doi.org/10.48550/ARXIV.2011.09694}\BibitemShut {NoStop}%
\bibitem [{\citenamefont {Baker}\ \emph {et~al.}(2023)\citenamefont {Baker},
  \citenamefont {Park}, \citenamefont {Yu}, \citenamefont {Ghukasyan},
  \citenamefont {Goktas},\ and\ \citenamefont {Radha}}]{Baker2023}%
  \BibitemOpen
  \bibfield  {author} {\bibinfo {author} {\bibfnamefont {J.~S.}\ \bibnamefont
  {Baker}}, \bibinfo {author} {\bibfnamefont {G.}~\bibnamefont {Park}},
  \bibinfo {author} {\bibfnamefont {K.}~\bibnamefont {Yu}}, \bibinfo {author}
  {\bibfnamefont {A.}~\bibnamefont {Ghukasyan}}, \bibinfo {author}
  {\bibfnamefont {O.}~\bibnamefont {Goktas}}, \ and\ \bibinfo {author}
  {\bibfnamefont {S.~K.}\ \bibnamefont {Radha}},\ }\href {\doibase
  https://doi.org/10.48550/arXiv.2305.05881} {\bibfield  {journal} {\bibinfo
  {journal} {arXiv preprint arXiv:2305.05881}\ } (\bibinfo {year} {2023}),\
  https://doi.org/10.48550/arXiv.2305.05881},\ \Eprint
  {http://arxiv.org/abs/2305.05881} {arXiv:2305.05881 [quant-ph]} \BibitemShut
  {NoStop}%
\bibitem [{\citenamefont {Sch{\'o}lkopf}\ \emph {et~al.}(2002)\citenamefont
  {Sch{\'o}lkopf}, \citenamefont {Smola}, \citenamefont {Bach} \emph
  {et~al.}}]{Scholkopf2002}%
  \BibitemOpen
  \bibfield  {author} {\bibinfo {author} {\bibfnamefont {B.}~\bibnamefont
  {Sch{\'o}lkopf}}, \bibinfo {author} {\bibfnamefont {A.~J.}\ \bibnamefont
  {Smola}}, \bibinfo {author} {\bibfnamefont {F.}~\bibnamefont {Bach}},  \emph
  {et~al.},\ }\href
  {http://mitpress.mit.edu/9780262194754/learning-with-kernels} {\emph
  {\bibinfo {title} {Learning with Kernels: Support Vector Machines,
  Regularization, Optimization, and Beyond}}}\ (\bibinfo  {publisher} {MIT
  Press},\ \bibinfo {year} {2002})\BibitemShut {NoStop}%
\bibitem [{\citenamefont {Schuld}\ and\ \citenamefont
  {Petruccione}(2018)}]{Schuld2018}%
  \BibitemOpen
  \bibfield  {author} {\bibinfo {author} {\bibfnamefont {M.}~\bibnamefont
  {Schuld}}\ and\ \bibinfo {author} {\bibfnamefont {F.}~\bibnamefont
  {Petruccione}},\ }\href
  {https://link.springer.com/book/10.1007/978-3-319-96424-9} {\emph {\bibinfo
  {title} {Supervised learning with quantum computers}}}\ (\bibinfo
  {publisher} {Springer},\ \bibinfo {year} {2018})\BibitemShut {NoStop}%
\bibitem [{\citenamefont {Schuld}(2021)}]{Schuld2021}%
  \BibitemOpen
  \bibfield  {author} {\bibinfo {author} {\bibfnamefont {M.}~\bibnamefont
  {Schuld}},\ }\href {\doibase https://doi.org/10.48550/arXiv.2101.11020}
  {\bibfield  {journal} {\bibinfo  {journal} {arXiv preprint arXiv:2101.11020}\
  } (\bibinfo {year} {2021}),\
  https://doi.org/10.48550/arXiv.2101.11020}\BibitemShut {NoStop}%
\bibitem [{\citenamefont {Harrow}\ \emph {et~al.}(2009)\citenamefont {Harrow},
  \citenamefont {Hassidim},\ and\ \citenamefont {Lloyd}}]{Harrow2009}%
  \BibitemOpen
  \bibfield  {author} {\bibinfo {author} {\bibfnamefont {A.~W.}\ \bibnamefont
  {Harrow}}, \bibinfo {author} {\bibfnamefont {A.}~\bibnamefont {Hassidim}}, \
  and\ \bibinfo {author} {\bibfnamefont {S.}~\bibnamefont {Lloyd}},\ }\href
  {\doibase https://doi.org/10.1103/PhysRevLett.103.150502} {\bibfield
  {journal} {\bibinfo  {journal} {Phys. Rev. Lett.}\ }\textbf {\bibinfo
  {volume} {103}},\ \bibinfo {pages} {150502} (\bibinfo {year}
  {2009})}\BibitemShut {NoStop}%
\bibitem [{\citenamefont {Havl{\'\i}{\v{c}}ek}\ \emph
  {et~al.}(2019)\citenamefont {Havl{\'\i}{\v{c}}ek}, \citenamefont
  {C{\'o}rcoles}, \citenamefont {Temme}, \citenamefont {Harrow}, \citenamefont
  {Kandala}, \citenamefont {Chow},\ and\ \citenamefont
  {Gambetta}}]{Havlivcek2019}%
  \BibitemOpen
  \bibfield  {author} {\bibinfo {author} {\bibfnamefont {V.}~\bibnamefont
  {Havl{\'\i}{\v{c}}ek}}, \bibinfo {author} {\bibfnamefont {A.~D.}\
  \bibnamefont {C{\'o}rcoles}}, \bibinfo {author} {\bibfnamefont
  {K.}~\bibnamefont {Temme}}, \bibinfo {author} {\bibfnamefont {A.~W.}\
  \bibnamefont {Harrow}}, \bibinfo {author} {\bibfnamefont {A.}~\bibnamefont
  {Kandala}}, \bibinfo {author} {\bibfnamefont {J.~M.}\ \bibnamefont {Chow}}, \
  and\ \bibinfo {author} {\bibfnamefont {J.~M.}\ \bibnamefont {Gambetta}},\
  }\href {\doibase https://doi.org/10.1038/s41586-019-0980-2} {\bibfield
  {journal} {\bibinfo  {journal} {Nature}\ }\textbf {\bibinfo {volume} {567}},\
  \bibinfo {pages} {209} (\bibinfo {year} {2019})}\BibitemShut {NoStop}%
\bibitem [{\citenamefont {Farhi}\ \emph {et~al.}(2014)\citenamefont {Farhi},
  \citenamefont {Goldstone},\ and\ \citenamefont {Gutmann}}]{Farhi2014}%
  \BibitemOpen
  \bibfield  {author} {\bibinfo {author} {\bibfnamefont {E.}~\bibnamefont
  {Farhi}}, \bibinfo {author} {\bibfnamefont {J.}~\bibnamefont {Goldstone}}, \
  and\ \bibinfo {author} {\bibfnamefont {S.}~\bibnamefont {Gutmann}},\ }\href
  {\doibase https://doi.org/10.48550/arXiv.1411.4028} {\bibfield  {journal}
  {\bibinfo  {journal} {arXiv preprint arXiv:1411.4028}\ } (\bibinfo {year}
  {2014}),\ https://doi.org/10.48550/arXiv.1411.4028}\BibitemShut {NoStop}%
\bibitem [{\citenamefont {Lloyd}\ \emph {et~al.}(2020)\citenamefont {Lloyd},
  \citenamefont {Schuld}, \citenamefont {Ijaz}, \citenamefont {Izaac},\ and\
  \citenamefont {Killoran}}]{Lloyd2020quantum}%
  \BibitemOpen
  \bibfield  {author} {\bibinfo {author} {\bibfnamefont {S.}~\bibnamefont
  {Lloyd}}, \bibinfo {author} {\bibfnamefont {M.}~\bibnamefont {Schuld}},
  \bibinfo {author} {\bibfnamefont {A.}~\bibnamefont {Ijaz}}, \bibinfo {author}
  {\bibfnamefont {J.}~\bibnamefont {Izaac}}, \ and\ \bibinfo {author}
  {\bibfnamefont {N.}~\bibnamefont {Killoran}},\ }\href {\doibase
  https://doi.org/10.48550/arXiv.2001.03622} {\bibfield  {journal} {\bibinfo
  {journal} {arXiv preprint arXiv:2001.03622}\ } (\bibinfo {year} {2020}),\
  https://doi.org/10.48550/arXiv.2001.03622}\BibitemShut {NoStop}%
\bibitem [{\citenamefont {Cincio}\ \emph {et~al.}(2018)\citenamefont {Cincio},
  \citenamefont {Suba\c{s}i}, \citenamefont {Sornborger},\ and\ \citenamefont
  {Coles}}]{Cincio2018}%
  \BibitemOpen
  \bibfield  {author} {\bibinfo {author} {\bibfnamefont {L.}~\bibnamefont
  {Cincio}}, \bibinfo {author} {\bibfnamefont {Y.}~\bibnamefont {Suba\c{s}i}},
  \bibinfo {author} {\bibfnamefont {A.~T.}\ \bibnamefont {Sornborger}}, \ and\
  \bibinfo {author} {\bibfnamefont {P.~J.}\ \bibnamefont {Coles}},\ }\href
  {\doibase https://doi.org/10.1088/1367-2630/aae94a} {\bibfield  {journal}
  {\bibinfo  {journal} {New J. Phys.}\ }\textbf {\bibinfo {volume} {20}},\
  \bibinfo {pages} {113022} (\bibinfo {year} {2018})}\BibitemShut {NoStop}%
\bibitem [{\citenamefont {Fanizza}\ \emph {et~al.}(2020)\citenamefont
  {Fanizza}, \citenamefont {Rosati}, \citenamefont {Skotiniotis}, \citenamefont
  {Calsamiglia},\ and\ \citenamefont {Giovannetti}}]{Fanizza2020}%
  \BibitemOpen
  \bibfield  {author} {\bibinfo {author} {\bibfnamefont {M.}~\bibnamefont
  {Fanizza}}, \bibinfo {author} {\bibfnamefont {M.}~\bibnamefont {Rosati}},
  \bibinfo {author} {\bibfnamefont {M.}~\bibnamefont {Skotiniotis}}, \bibinfo
  {author} {\bibfnamefont {J.}~\bibnamefont {Calsamiglia}}, \ and\ \bibinfo
  {author} {\bibfnamefont {V.}~\bibnamefont {Giovannetti}},\ }\href {\doibase
  https://doi.org/10.1103/PhysRevLett.124.060503} {\bibfield  {journal}
  {\bibinfo  {journal} {Phys. Rev. Lett.}\ }\textbf {\bibinfo {volume} {124}},\
  \bibinfo {pages} {060503} (\bibinfo {year} {2020})}\BibitemShut {NoStop}%
\bibitem [{\citenamefont {Huang}\ \emph {et~al.}(2022)\citenamefont {Huang},
  \citenamefont {Keung},\ and\ \citenamefont {Preskill}}]{Huang2020}%
  \BibitemOpen
  \bibfield  {author} {\bibinfo {author} {\bibfnamefont {H.-Y.}\ \bibnamefont
  {Huang}}, \bibinfo {author} {\bibfnamefont {R.}~\bibnamefont {Keung}}, \ and\
  \bibinfo {author} {\bibfnamefont {J.}~\bibnamefont {Preskill}},\ }\href
  {\doibase https://doi.org/10.1038/s41567-020-0932-7} {\bibfield  {journal}
  {\bibinfo  {journal} {Nat. Phys.}\ }\textbf {\bibinfo {volume} {16}},\
  \bibinfo {pages} {1050} (\bibinfo {year} {2022})}\BibitemShut {NoStop}%
\bibitem [{\citenamefont {Brouard}\ \emph {et~al.}(2022)\citenamefont
  {Brouard}, \citenamefont {Mariette}, \citenamefont {Flamary},\ and\
  \citenamefont {Vialaneix}}]{Brouard2022}%
  \BibitemOpen
  \bibfield  {author} {\bibinfo {author} {\bibfnamefont {C.}~\bibnamefont
  {Brouard}}, \bibinfo {author} {\bibfnamefont {J.}~\bibnamefont {Mariette}},
  \bibinfo {author} {\bibfnamefont {R.}~\bibnamefont {Flamary}}, \ and\
  \bibinfo {author} {\bibfnamefont {N.}~\bibnamefont {Vialaneix}},\ }\href
  {\doibase https://doi.org/10.1093/nargab/lqac014} {\bibfield  {journal}
  {\bibinfo  {journal} {NAR Genom. Bioinform.}\ }\textbf {\bibinfo {volume}
  {4}} (\bibinfo {year} {2022}),\
  https://doi.org/10.1093/nargab/lqac014}\BibitemShut {NoStop}%
\bibitem [{\citenamefont {Xue}\ \emph {et~al.}(2020)\citenamefont {Xue},
  \citenamefont {Song},\ and\ \citenamefont {Xu}}]{Xue2020}%
  \BibitemOpen
  \bibfield  {author} {\bibinfo {author} {\bibfnamefont {H.}~\bibnamefont
  {Xue}}, \bibinfo {author} {\bibfnamefont {Y.}~\bibnamefont {Song}}, \ and\
  \bibinfo {author} {\bibfnamefont {H.-M.}\ \bibnamefont {Xu}},\ }\href
  {\doibase https://doi.org/10.1016/j.knosys.2019.105272} {\bibfield  {journal}
  {\bibinfo  {journal} {Knowl. Based Syst.}\ }\textbf {\bibinfo {volume}
  {191}},\ \bibinfo {pages} {105272} (\bibinfo {year} {2020})}\BibitemShut
  {NoStop}%
\bibitem [{\citenamefont {Gautam}\ \emph {et~al.}(2019)\citenamefont {Gautam},
  \citenamefont {Balaji}, \citenamefont {K.}, \citenamefont {Tiwari},\ and\
  \citenamefont {Ahuja}}]{Gautam2019}%
  \BibitemOpen
  \bibfield  {author} {\bibinfo {author} {\bibfnamefont {C.}~\bibnamefont
  {Gautam}}, \bibinfo {author} {\bibfnamefont {R.}~\bibnamefont {Balaji}},
  \bibinfo {author} {\bibfnamefont {S.}~\bibnamefont {K.}}, \bibinfo {author}
  {\bibfnamefont {A.}~\bibnamefont {Tiwari}}, \ and\ \bibinfo {author}
  {\bibfnamefont {K.}~\bibnamefont {Ahuja}},\ }\href {\doibase
  https://doi.org/10.1016/j.knosys.2018.11.030} {\bibfield  {journal} {\bibinfo
   {journal} {Knowl. Based Syst.}\ }\textbf {\bibinfo {volume} {165}},\
  \bibinfo {pages} {241} (\bibinfo {year} {2019})}\BibitemShut {NoStop}%
\bibitem [{\citenamefont {Steinward}\ and\ \citenamefont
  {Christmann}(2008)}]{Steinwart2008}%
  \BibitemOpen
  \bibfield  {author} {\bibinfo {author} {\bibfnamefont {I.}~\bibnamefont
  {Steinward}}\ and\ \bibinfo {author} {\bibfnamefont {A.}~\bibnamefont
  {Christmann}},\ }\href {\doibase https://doi.org/10.1007/978-0-387-77242-4}
  {\emph {\bibinfo {title} {Support Vector Machines}}}\ (\bibinfo  {publisher}
  {Springer Science \& Business Media},\ \bibinfo {year} {2008})\BibitemShut
  {NoStop}%
\bibitem [{\citenamefont {Kingma}\ and\ \citenamefont
  {Ba}(2014)}]{kingma2014adam}%
  \BibitemOpen
  \bibfield  {author} {\bibinfo {author} {\bibfnamefont {D.~P.}\ \bibnamefont
  {Kingma}}\ and\ \bibinfo {author} {\bibfnamefont {J.}~\bibnamefont {Ba}},\
  }\href {\doibase https://doi.org/10.48550/arXiv.1412.6980} {\bibfield
  {journal} {\bibinfo  {journal} {arXiv preprint arXiv:1412.6980}\ } (\bibinfo
  {year} {2014}),\ https://doi.org/10.48550/arXiv.1412.6980}\BibitemShut
  {NoStop}%
\bibitem [{\citenamefont {Pedregosa}\ \emph {et~al.}(2011)\citenamefont
  {Pedregosa}, \citenamefont {Varoquaux}, \citenamefont {Gramfort},
  \citenamefont {Michel}, \citenamefont {Thirion}, \citenamefont {Grisel},
  \citenamefont {Blondel}, \citenamefont {Prettenhofer}, \citenamefont {Weiss},
  \citenamefont {Dubourg}, \citenamefont {Vanderplas}, \citenamefont {Passos},
  \citenamefont {Cournapeau}, \citenamefont {Brucher}, \citenamefont {Perrot},\
  and\ \citenamefont {Duchesnay}}]{scikit-learn}%
  \BibitemOpen
  \bibfield  {author} {\bibinfo {author} {\bibfnamefont {F.}~\bibnamefont
  {Pedregosa}}, \bibinfo {author} {\bibfnamefont {G.}~\bibnamefont
  {Varoquaux}}, \bibinfo {author} {\bibfnamefont {A.}~\bibnamefont {Gramfort}},
  \bibinfo {author} {\bibfnamefont {V.}~\bibnamefont {Michel}}, \bibinfo
  {author} {\bibfnamefont {B.}~\bibnamefont {Thirion}}, \bibinfo {author}
  {\bibfnamefont {O.}~\bibnamefont {Grisel}}, \bibinfo {author} {\bibfnamefont
  {M.}~\bibnamefont {Blondel}}, \bibinfo {author} {\bibfnamefont
  {P.}~\bibnamefont {Prettenhofer}}, \bibinfo {author} {\bibfnamefont
  {R.}~\bibnamefont {Weiss}}, \bibinfo {author} {\bibfnamefont
  {V.}~\bibnamefont {Dubourg}}, \bibinfo {author} {\bibfnamefont
  {J.}~\bibnamefont {Vanderplas}}, \bibinfo {author} {\bibfnamefont
  {A.}~\bibnamefont {Passos}}, \bibinfo {author} {\bibfnamefont
  {D.}~\bibnamefont {Cournapeau}}, \bibinfo {author} {\bibfnamefont
  {M.}~\bibnamefont {Brucher}}, \bibinfo {author} {\bibfnamefont
  {M.}~\bibnamefont {Perrot}}, \ and\ \bibinfo {author} {\bibfnamefont
  {E.}~\bibnamefont {Duchesnay}},\ }\href
  {https://jmlr.org/papers/v12/pedregosa11a.html} {\bibfield  {journal}
  {\bibinfo  {journal} {J. Mach. Learn. Res.}\ }\textbf {\bibinfo {volume}
  {12}},\ \bibinfo {pages} {2825} (\bibinfo {year} {2011})}\BibitemShut
  {NoStop}%
\bibitem [{\citenamefont {Steinwart}\ \emph {et~al.}(2006)\citenamefont
  {Steinwart}, \citenamefont {Hush},\ and\ \citenamefont
  {Scovel}}]{Steinwart2006}%
  \BibitemOpen
  \bibfield  {author} {\bibinfo {author} {\bibfnamefont {I.}~\bibnamefont
  {Steinwart}}, \bibinfo {author} {\bibfnamefont {D.}~\bibnamefont {Hush}}, \
  and\ \bibinfo {author} {\bibfnamefont {C.}~\bibnamefont {Scovel}},\ }\href
  {\doibase https://doi.org/10.1109/TIT.2006.881713} {\bibfield  {journal}
  {\bibinfo  {journal} {IEEE Trans. Inf.}\ }\textbf {\bibinfo {volume} {52}},\
  \bibinfo {pages} {4635} (\bibinfo {year} {2006})}\BibitemShut {NoStop}%
\bibitem [{\citenamefont {Micchelli}\ \emph {et~al.}(2006)\citenamefont
  {Micchelli}, \citenamefont {Xu},\ and\ \citenamefont
  {Zhang}}]{Micchelli2006}%
  \BibitemOpen
  \bibfield  {author} {\bibinfo {author} {\bibfnamefont {C.~A.}\ \bibnamefont
  {Micchelli}}, \bibinfo {author} {\bibfnamefont {Y.}~\bibnamefont {Xu}}, \
  and\ \bibinfo {author} {\bibfnamefont {H.}~\bibnamefont {Zhang}},\ }\href
  {http://jmlr.org/papers/v7/micchelli06a.html} {\bibfield  {journal} {\bibinfo
   {journal} {J. Mach. Learn. Res.}\ }\textbf {\bibinfo {volume} {7}},\
  \bibinfo {pages} {2667} (\bibinfo {year} {2006})}\BibitemShut {NoStop}%
\bibitem [{\citenamefont {Diamond}\ and\ \citenamefont
  {Boyd}(2016)}]{cvxpy2016}%
  \BibitemOpen
  \bibfield  {author} {\bibinfo {author} {\bibfnamefont {S.}~\bibnamefont
  {Diamond}}\ and\ \bibinfo {author} {\bibfnamefont {S.}~\bibnamefont {Boyd}},\
  }\href {https://www.jmlr.org/papers/volume17/15-408/15-408.pdf} {\bibfield
  {journal} {\bibinfo  {journal} {J. Mach. Learn. Res.}\ }\textbf {\bibinfo
  {volume} {17}},\ \bibinfo {pages} {1} (\bibinfo {year} {2016})}\BibitemShut
  {NoStop}%
\bibitem [{\citenamefont {Agrawal}\ \emph
  {et~al.}(2019{\natexlab{a}})\citenamefont {Agrawal}, \citenamefont
  {Verschueren}, \citenamefont {Diamond},\ and\ \citenamefont
  {Boyd}}]{cvxpy2018}%
  \BibitemOpen
  \bibfield  {author} {\bibinfo {author} {\bibfnamefont {A.}~\bibnamefont
  {Agrawal}}, \bibinfo {author} {\bibfnamefont {R.}~\bibnamefont
  {Verschueren}}, \bibinfo {author} {\bibfnamefont {S.}~\bibnamefont
  {Diamond}}, \ and\ \bibinfo {author} {\bibfnamefont {S.}~\bibnamefont
  {Boyd}},\ }\href {\doibase https://doi.org/10.48550/arXiv.1709.04494}
  {\bibfield  {journal} {\bibinfo  {journal} {J. Control. Decis.}\ } (\bibinfo
  {year} {2019}{\natexlab{a}}),\ https://doi.org/10.48550/arXiv.1709.04494},\
  \Eprint {http://arxiv.org/abs/1709.04494} {arXiv:1709.04494 [math.OC]}
  \BibitemShut {NoStop}%
\bibitem [{\citenamefont {O'Donoghue}\ \emph {et~al.}(2016)\citenamefont
  {O'Donoghue}, \citenamefont {Chu}, \citenamefont {Parikh},\ and\
  \citenamefont {Boyd}}]{odonoghue2016}%
  \BibitemOpen
  \bibfield  {author} {\bibinfo {author} {\bibfnamefont {B.}~\bibnamefont
  {O'Donoghue}}, \bibinfo {author} {\bibfnamefont {E.}~\bibnamefont {Chu}},
  \bibinfo {author} {\bibfnamefont {N.}~\bibnamefont {Parikh}}, \ and\ \bibinfo
  {author} {\bibfnamefont {S.}~\bibnamefont {Boyd}},\ }\href {\doibase
  https://doi.org/10.1007/s10957-016-0892-3} {\bibfield  {journal} {\bibinfo
  {journal} {Journal of Optimization Theory and Applications}\ }\textbf
  {\bibinfo {volume} {169}},\ \bibinfo {pages} {1042} (\bibinfo {year}
  {2016})}\BibitemShut {NoStop}%
\bibitem [{\citenamefont {Agrawal}\ \emph
  {et~al.}(2019{\natexlab{b}})\citenamefont {Agrawal}, \citenamefont {Amos},
  \citenamefont {Barratt}, \citenamefont {Boyd}, \citenamefont {Diamond},\ and\
  \citenamefont {Kolter}}]{cvxpylayers2019}%
  \BibitemOpen
  \bibfield  {author} {\bibinfo {author} {\bibfnamefont {A.}~\bibnamefont
  {Agrawal}}, \bibinfo {author} {\bibfnamefont {B.}~\bibnamefont {Amos}},
  \bibinfo {author} {\bibfnamefont {S.}~\bibnamefont {Barratt}}, \bibinfo
  {author} {\bibfnamefont {S.}~\bibnamefont {Boyd}}, \bibinfo {author}
  {\bibfnamefont {S.}~\bibnamefont {Diamond}}, \ and\ \bibinfo {author}
  {\bibfnamefont {Z.}~\bibnamefont {Kolter}},\ }in\ \href
  {https://proceedings.neurips.cc/paper/2019/hash/9ce3c52fc54362e22053399d3181c638-Abstract.html}
  {\emph {\bibinfo {booktitle} {Advances in Neural Information Processing
  Systems}}}\ (\bibinfo {year} {2019})\BibitemShut {NoStop}%
\bibitem [{\citenamefont {Agrawal}\ \emph
  {et~al.}(2019{\natexlab{c}})\citenamefont {Agrawal}, \citenamefont {Barratt},
  \citenamefont {Boyd}, \citenamefont {Busseti},\ and\ \citenamefont
  {Moursi}}]{diffcp2019}%
  \BibitemOpen
  \bibfield  {author} {\bibinfo {author} {\bibfnamefont {A.}~\bibnamefont
  {Agrawal}}, \bibinfo {author} {\bibfnamefont {S.}~\bibnamefont {Barratt}},
  \bibinfo {author} {\bibfnamefont {S.}~\bibnamefont {Boyd}}, \bibinfo {author}
  {\bibfnamefont {E.}~\bibnamefont {Busseti}}, \ and\ \bibinfo {author}
  {\bibfnamefont {W.}~\bibnamefont {Moursi}},\ }\href {\doibase
  https://doi.org/10.48550/arXiv.1904.09043} {\bibfield  {journal} {\bibinfo
  {journal} {Journal of Applied and Numerical Optimization}\ }\textbf {\bibinfo
  {volume} {1}},\ \bibinfo {pages} {107} (\bibinfo {year}
  {2019}{\natexlab{c}})}\BibitemShut {NoStop}%
\bibitem [{\citenamefont {Bergholm}\ \emph {et~al.}(2022)\citenamefont
  {Bergholm}, \citenamefont {Izaac}, \citenamefont {Schuld}, \citenamefont
  {Gogolin}, \citenamefont {Ahmed} \emph {et~al.}}]{pennylane2022}%
  \BibitemOpen
  \bibfield  {author} {\bibinfo {author} {\bibfnamefont {V.}~\bibnamefont
  {Bergholm}}, \bibinfo {author} {\bibfnamefont {J.}~\bibnamefont {Izaac}},
  \bibinfo {author} {\bibfnamefont {M.}~\bibnamefont {Schuld}}, \bibinfo
  {author} {\bibfnamefont {C.}~\bibnamefont {Gogolin}}, \bibinfo {author}
  {\bibfnamefont {S.}~\bibnamefont {Ahmed}},  \emph {et~al.},\ }\href {\doibase
  https://doi.org/10.48550/arXiv.1811.04968} {\enquote {\bibinfo {title}
  {Pennylane: Automatic differentiation of hybrid quantum-classical
  computations},}\ } (\bibinfo {year} {2022}),\ \Eprint
  {http://arxiv.org/abs/1811.04968} {arXiv:1811.04968 [quant-ph]} \BibitemShut
  {NoStop}%
\bibitem [{\citenamefont {Paszke}\ \emph {et~al.}(2019)\citenamefont {Paszke},
  \citenamefont {Gross}, \citenamefont {Massa}, \citenamefont {Lerer},
  \citenamefont {Bradbury}, \citenamefont {Chanan}, \citenamefont {Killeen},
  \citenamefont {Lin}, \citenamefont {Gimelshein}, \citenamefont {Antiga} \emph
  {et~al.}}]{pytorch2019}%
  \BibitemOpen
  \bibfield  {author} {\bibinfo {author} {\bibfnamefont {A.}~\bibnamefont
  {Paszke}}, \bibinfo {author} {\bibfnamefont {S.}~\bibnamefont {Gross}},
  \bibinfo {author} {\bibfnamefont {F.}~\bibnamefont {Massa}}, \bibinfo
  {author} {\bibfnamefont {A.}~\bibnamefont {Lerer}}, \bibinfo {author}
  {\bibfnamefont {J.}~\bibnamefont {Bradbury}}, \bibinfo {author}
  {\bibfnamefont {G.}~\bibnamefont {Chanan}}, \bibinfo {author} {\bibfnamefont
  {T.}~\bibnamefont {Killeen}}, \bibinfo {author} {\bibfnamefont
  {Z.}~\bibnamefont {Lin}}, \bibinfo {author} {\bibfnamefont {N.}~\bibnamefont
  {Gimelshein}}, \bibinfo {author} {\bibfnamefont {L.}~\bibnamefont {Antiga}},
  \emph {et~al.},\ }\href {\doibase https://doi.org/10.48550/arXiv.1912.01703}
  {\bibfield  {journal} {\bibinfo  {journal} {Advances in neural information
  processing systems}\ }\textbf {\bibinfo {volume} {32}} (\bibinfo {year}
  {2019}),\ https://doi.org/10.48550/arXiv.1912.01703}\BibitemShut {NoStop}%
\bibitem [{Note1()}]{Note1}%
  \BibitemOpen
  \bibinfo {note} {Covalent: \protect \url
  {https://www.covalent.xyz}}\BibitemShut {NoStop}%
\bibitem [{Note2()}]{Note2}%
  \BibitemOpen
  \bibinfo {note} {\protect \texttt {make\protect \_classification}: \protect
  \url
  {https://scikit-learn.org/stable/modules/generated/sklearn.datasets.make_classification.html}}\BibitemShut
  {NoStop}%
\bibitem [{\citenamefont {Guyon}(2003)}]{Guyon2003}%
  \BibitemOpen
  \bibfield  {author} {\bibinfo {author} {\bibfnamefont {I.}~\bibnamefont
  {Guyon}},\ }in\ \href
  {http://clopinet.com/isabelle/Projects/NIPS2003/Slides/NIPS2003-Datasets.pdf}
  {\emph {\bibinfo {booktitle} {NIPS 2003 workshop on feature extraction and
  feature selection}}},\ Vol.\ \bibinfo {volume} {253}\ (\bibinfo {year}
  {2003})\ p.~\bibinfo {pages} {40}\BibitemShut {NoStop}%
\bibitem [{Note3()}]{Note3}%
  \BibitemOpen
  \bibinfo {note} {\protect \texttt {MinMaxScaler}: \protect \url
  {https://scikit-learn.org/stable/modules/generated/sklearn.preprocessing.MinMaxScaler.html}}\BibitemShut
  {NoStop}%
\bibitem [{Note4()}]{Note4}%
  \BibitemOpen
  \bibinfo {note} {\protect \texttt {SVC}: \protect \url
  {https://scikit-learn.org/stable/modules/generated/sklearn.svm.SVC.html}}\BibitemShut
  {NoStop}%
\bibitem [{Note5()}]{Note5}%
  \BibitemOpen
  \bibinfo {note} {See Supplementary Information [publisher url] for examples
  of two-dimensional datasets.}\BibitemShut {Stop}%
\bibitem [{\citenamefont {Jerbi}\ \emph {et~al.}(2023)\citenamefont {Jerbi},
  \citenamefont {Fiderer}, \citenamefont {Poulsen~Nautrup}, \citenamefont
  {K{\"u}bler}, \citenamefont {Briegel},\ and\ \citenamefont
  {Dunjko}}]{jerbi2023quantum}%
  \BibitemOpen
  \bibfield  {author} {\bibinfo {author} {\bibfnamefont {S.}~\bibnamefont
  {Jerbi}}, \bibinfo {author} {\bibfnamefont {L.~J.}\ \bibnamefont {Fiderer}},
  \bibinfo {author} {\bibfnamefont {H.}~\bibnamefont {Poulsen~Nautrup}},
  \bibinfo {author} {\bibfnamefont {J.~M.}\ \bibnamefont {K{\"u}bler}},
  \bibinfo {author} {\bibfnamefont {H.~J.}\ \bibnamefont {Briegel}}, \ and\
  \bibinfo {author} {\bibfnamefont {V.}~\bibnamefont {Dunjko}},\ }\href
  {\doibase https://doi.org/10.1038/s41467-023-36159-y} {\bibfield  {journal}
  {\bibinfo  {journal} {Nature Communications}\ }\textbf {\bibinfo {volume}
  {14}},\ \bibinfo {pages} {517} (\bibinfo {year} {2023})}\BibitemShut
  {NoStop}%
\bibitem [{\citenamefont {Glick}\ \emph {et~al.}(2021)\citenamefont {Glick},
  \citenamefont {Gujarati}, \citenamefont {Corcoles}, \citenamefont {Kim},
  \citenamefont {Kandala}, \citenamefont {Gambetta},\ and\ \citenamefont
  {Temme}}]{glick2021covariant}%
  \BibitemOpen
  \bibfield  {author} {\bibinfo {author} {\bibfnamefont {J.~R.}\ \bibnamefont
  {Glick}}, \bibinfo {author} {\bibfnamefont {T.~P.}\ \bibnamefont {Gujarati}},
  \bibinfo {author} {\bibfnamefont {A.~D.}\ \bibnamefont {Corcoles}}, \bibinfo
  {author} {\bibfnamefont {Y.}~\bibnamefont {Kim}}, \bibinfo {author}
  {\bibfnamefont {A.}~\bibnamefont {Kandala}}, \bibinfo {author} {\bibfnamefont
  {J.~M.}\ \bibnamefont {Gambetta}}, \ and\ \bibinfo {author} {\bibfnamefont
  {K.}~\bibnamefont {Temme}},\ }\href {\doibase
  https://doi.org/10.48550/arXiv.2105.03406} {\bibfield  {journal} {\bibinfo
  {journal} {arXiv preprint arXiv:2105.03406}\ } (\bibinfo {year} {2021}),\
  https://doi.org/10.48550/arXiv.2105.03406}\BibitemShut {NoStop}%
\end{thebibliography}%

\clearpage
\onecolumngrid
\section*{Supplementary Information:\\Quantum-Classical Multiple Kernel Learning \label{sec:supplementary}}
\setcounter{figure}{0}
\renewcommand\thefigure{S\arabic{figure}}   

\begin{figure}[h]
    \centering
    \includegraphics[width=0.85\linewidth]{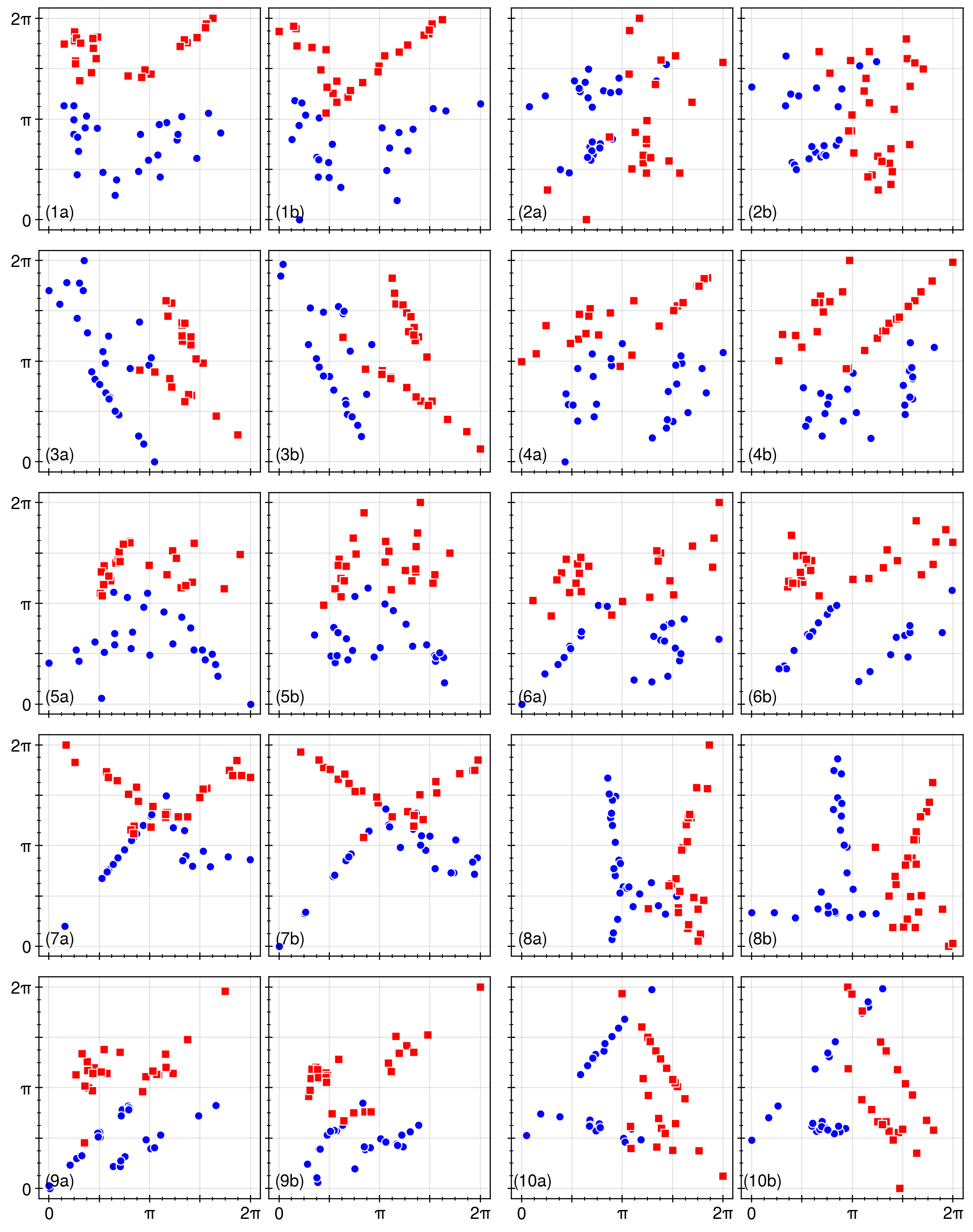}
    \caption{Ten examples of $d=2$ dimensional classification datasets corresponding to Sec. \ref{subsec:dataset-preparation}. Each pair of plots shows the equally-split training and testing subsets (left (a) and right (b), respectively). All horizontal and vertical axes range from 0 to $2\pi$. Square scatter points (red) belong one class and round scatter points (blue) belong to the other.}
    \label{fig:SM-visualize-2d}
\end{figure}

\end{document}